%% file: paper.tex
\documentclass[fleqn,usenatbib]{mnras}

\usepackage{newtxtext,newtxmath}

\usepackage[T1]{fontenc}
\usepackage{ae,aecompl}

\usepackage{graphicx}	
\usepackage{amsmath}	
\usepackage[usenames,dvipsnames]{xcolor}
\usepackage{longtable}
\usepackage{pdfpages}

\newcommand{\Teff}{\mbox{$T_{\rm eff}$}}

\newcommand{\kepler}{{\em Kepler\/}}
\newcommand{\tess}{{\em TESS\/}}
\newcommand{\bcep}{\mbox{$\beta$~Cep}}
\newcommand{\dsct}{\mbox{$\delta$~Sct}}

\newcommand{\cd}{\mbox{d$^{-1}$}}
\newcommand{\Dnu}{\mbox{$\Delta\nu$}}
\newcommand{\kms}{\mbox{km\,s$^{-1}$}}
\newcommand{\vsini}{\mbox{$v\sin i$}}
\newcommand{\Lsun}{\mbox{L$_{\odot}$}}
\newcommand{\Msun}{\mbox{M$_{\odot}$}}

\newcommand{\echelle}{{\'e}chelle}

\newcommand{\firstnew}[1]{#1} 
\newcommand{\new}[1]{{\color{red}\textbf{#1}}}
\renewcommand{\new}[1]{#1} 

\title[Pulsating B stars in Sco--Cen]{Pulsating B stars in the \firstnew{Scorpius}--Centaurus Association with \tess}

\author[Sharma et al.]{%
Awshesh N. Sharma$^{1,2}$\thanks{E-mail: asharma@es.iitr.ac.in}, 
Timothy R. Bedding$^{2,3}$\thanks{E-mail: tim.bedding@sydney.edu.au},
Hideyuki Saio$^4$ and \and
Timothy R. White$^{2,3}$
\smallskip
\\
$^1$Department of Earth Sciences, Indian Institute of Technology Roorkee, Roorkee, Uttarakhand 247667 India\\
$^2$Sydney Institute for Astronomy, School of Physics, University of Sydney 2006, Australia \\
$^3$Stellar Astrophysics Centre, Department of Physics and Astronomy, Aarhus University, 8000 Aarhus C, Denmark\\
$^4$Astronomical Institute, Graduate School of Science, Tohoku University, Sendai 980-8578, Japan
}

\date{}

\pubyear{2021}

\begin{document}
\label{firstpage}
\pagerange{\pageref{firstpage}--\pageref{lastpage}}
\maketitle

\begin{abstract}
We study 119 B stars located in the \firstnew{Scorpius}--Centaurus Association using data from NASA's \tess\ Mission.  We see pulsations in 81 stars (68\%) across the full range of effective temperatures.  In particular, we confirm previous reports of low-frequency pulsations in stars whose temperatures fall between the instability strips of SPB stars (slowly pulsating B stars) and $\delta$~Scuti stars.  By taking the stellar densities into account, we conclude that these cannot be p~modes and confirm previous suggestions that these are probably rapidly-rotating SPB stars. We also confirm that they follow two period--luminosity relations that are consistent with prograde sectoral g~modes that are dipole ($l=m=1$) and quadrupole ($l=m=2$), respectively.  One of the stars ($\xi^2$~Cen) is a hybrid pulsator that shows regular spacings in both g~and p~modes.  We \firstnew{confirm} that $\alpha$~Cru has low-amplitude p-mode pulsations, making it one of the brightest $\beta$~Cephei stars in the sky.  
We also find several interesting binaries, including a very short-period heartbeat star (HD~132094), 
a previously unknown eclipsing binary ($\pi$~Lup), and
an eclipsing binary with high-amplitude tidally driven pulsations (HR~5846).
The results clearly demonstrate the power of \tess\ for studying variability in stellar associations.
\end{abstract}

\begin{keywords}
stars: oscillations, 
\end{keywords}




\section{Introduction}

The \textit{Transiting Exoplanet Survey Satellite} (\tess; \citealt{Ricker++2015}) is providing high-precision photometry for bright stars across a large fraction of the sky, which allows a systematic study of pulsations in nearby stellar associations.  We have used \tess\ light curves to examine the variability of more than a hundred B-type stars in the \firstnew{Scorpius}--Centaurus Association (Sco--Cen; Sco~OB2).  This is the nearest OB association, with member stars having ages in the range 5--16\,Myr \citep{Blaauw1946,Jones+Shobbrook1974,Balona+Feast1975,deGeus++1989,deZeeuw++1999,Preibisch+Mamajek2008,Pecaut+Mamajek2016,Murphy2021, Wright++2022}. 

Pulsating B stars are generally divided into two classes.  At hotter effective temperatures (spectral types B0--B3) we have the \bcep{}hei stars, which pulsate in low-order pressure modes \citep[p modes;][]{Stankov+Handler2005,Thoul2009}.  At cooler temperatures (typically B3--B8) we find the Slowly Pulsating B-type stars (SPB stars), which pulsate in high-order gravity modes \citep[g modes;][]{Waelkens1991,DeCat++2011}.  Both the p and g modes are known to be excited by a heat-engine mechanism acting on the opacities of iron-group elements \citep{Dziembowski+Pamiatnykh1993,Gautschy+Saio1993,Dziembowski++1993}.

Research into pulsating B stars has been greatly helped by photometry from space missions such as MOST, WIRE, CoRoT, BRITE, \kepler/K2, and \tess\ (for reviews, see \citealt{DeCat++2011,Bowman2020}).  Current areas of research include understanding the driving mechanism in the low-metallicity stars of the Magellanic Clouds \citep{Salmon++2012}, testing whether opacities need to be modified \citep{Lenz2011,Mozdzierski++2019,Daszynska-Daszkiewicz++2017,Daszynska-Daszkiewicz++2020}, and measuring the amount of internal rotation and mixing \citep{Degroote++2010, Papics++2017, Pedersen++2021, Pedersen2022}.



Another topic of interest relates to the repeated reports of pulsations in stars that are cooler than expected for SPB stars, with effective temperatures falling between those of SPB stars and \dsct\ stars 
\citep{Handler++2007,Handler++2008,Majewska-Swierzbinowicz++2008,Degroote2009,Saesen++2010,Saesen++2013,Mowlavi2013,Mowlavi2016,Mozdzierski++2014,Lata++2014,Balona++2015,Balona++2016,Daszynska-Daszkiewicz++2017,Balona+Ozuyar2020,van-Heerden++2020}.  These have sometimes been called `Maia variables' but we avoid this term because they have not been established as a separate class of variables \citep[e.g.,][]{Aerts+Kolenberg2005}, and also because \kepler\ K2 photometry has shown that Maia itself (20~Tau = HD~23408 in the Pleiades) is not actually a member of the group \citep{White++2017}.

The large sky coverage of the \tess\ mission makes it ideally suited to studying B stars \citep[e.g.,][]{Balona++2019,Pedersen++2019,Burssens++2020}.
In this paper, we study variability in 119 B-type stars in Sco--Cen using data from \tess.  We focus on pulsations, but also report on eclipsing binaries and other types of variability.

\section{TESS Data} \label{sec:lightcurves}

Our sample of B-type stars in the Sco--Cen Association comes from \citet{Rizzuto++2011}, with a few additional stars that were considered members by \citet{Pecaut+Mamajek2016}. \tess\ covers the sky in partially overlapping sectors, with each sector being observed for about 27\,d (two orbits).  Note that \tess\ has not yet covered a band within 6\degr\ of the ecliptic plane that includes 56 B stars that lie in the Upper \firstnew{Scorpius} region of Sco--Cen\footnote{part of Scorpius was observed by \kepler\ K2 Mission and light curves for some bright stars have been made by \citet{Pope++2019} using halo photometry.}.  This leaves \new{109} B~stars with \tess\ data from Cycle~1, as listed in Table~\ref{tab:sample}.  Most were observed in one of Sectors~10, 11 or 12 (which covered from 26 March to 19 June 2019).  Note that 10 stars having observations in both Sectors 10 and~11.  
\new{The HD numbers for these stars are:
95324,
97583,
103079,
104900,
105937,
106490,
107696,
108257,
108483, and
109195.}
We also included \firstnew{10} B~stars that were missed in Cycle~1 (presumably because they landed in gaps between CCDs) but have data in Cycle~3 (Sectors 36, 37, 38 or 39). \new{The HD numbers for these stars are: 
90264,
93563, 
113902,
114529,
123445,
137432,
138221,
138923,
139365, and
150638.
}


The light curves and Fourier amplitude spectra for all stars are shown in the on-line supplementary material.
For most stars in our sample, 2-minute \tess\ light curves were available. We used the {\tt lightkurve} package \citep{lightkurve2018} to download the PDCSAP\footnote{Pre-search Data Conditioning Simple Aperture Photometry} light curves, which were calculated by the SPOC (Science Processing Operations Center). 
For 23 stars for which no 2-minute observations were available, we used light curves from the 30-minute full-frame images (FFIs) produced by the MIT Quick Look Pipeline \citep[QLP;][]{Huang++2020}.  


For $\beta$~Cen and $\alpha$~Cru, which are two of the brightest stars in the sample and are heavily saturated, we constructed 2-minute light curves ourselves from the pixel-level data. Due to the large size of the target pixel file of $\beta$~Cen, it was not processed by the TESS photometric pipeline. As a result, not only was there no available SPOC PDCSAP light curve, but the background had not been estimated and subtracted from the flux values in the target pixel file. We estimated the background using the {\tt tessbkgd}\footnote{\url{github.com/hvidy/tessbkgd}} python package, and calculated a simple aperture photometry light curve using {\tt lightkurve}.  For $\alpha$~Cru, a SPOC PDCSAP light curve was available but the selected pixels did not fully capture the bleed trails that result from saturation, and so aperture losses meant this light curve was of low quality. We instead generated a `halo' photometry light curve from the unsaturated pixels in the wings of the PSF using the {\tt halophot}\footnote{\url{github.com/hvidy/halophot}} package \citep{White++2017,Pope++2019}.  The light curves for $\alpha$~Cru and $\beta$~Cen are discussed in Secs.~\ref{sec:alpha-cru} and~\ref{sec:beta-cen}.


\section{Stellar properties}
\label{sec:stellar-properties}

The spectral types, effective temperatures and luminosities for our sample are listed in Table~\ref{tab:sample}.  
Most spectral types were taken from the five-volume University of Michigan Catalogue \citep{Houk+Cowley1975,Houk+Cowley1978,Houk1982,Houk+Smith-Moore1988,Houk+Swift1999}.
The exceptions were $\eta$~Cen, $\lambda$~Cru and $\mu$~Lup, which are known to be emission-line (Be) stars but are not classified as such in the Michigan Catalogue.  For these stars we adopted the spectral types measured by \citet{Levenhagen+Leister2006}.  


We estimated effective temperatures from spectral types using an updated version of Table~5 of \citet{Pecaut+Mamajek2013}\footnote{this list is maintained at \url{http://www.pas.rochester.edu/~emamajek/EEM_dwarf_UBVIJHK_colors_Teff.txt}}.
\firstnew{We note that there are no effective temperatures listed in the SIMBAD database for most of our stars.  \citet{Zorec+Royer2012} give temperatures from Str\"omgren photometry for about half of the coolest 50 stars in our sample, and these agree to better than 0.02 in $\log(\Teff)$ in most cases.} 
\new{We have not listed uncertainties for effective temperatures in Table~\ref{tab:sample} but we suggest a reasonable value is $\pm0.02$ in $\log(\Teff)$.}
\firstnew{An advantage of using} spectral types rather than photometry to estimate effective temperatures is that the former are likely to be more reliable for binary systems.

We estimated stellar luminosities from $V$ magnitudes and parallaxes, using bolometric corrections by \citet{Pecaut+Mamajek2013}.  \firstnew{For most stars we used Gaia EDR3 parallaxes \citep{Gaia++2020-EDR3}. We used Hipparcos parallaxes \citep{van-Leeuwen2007} for 16 stars that do not have Gaia EDR3 parallaxes (these are mainly the brightest stars in our sample), and for one star whose EDR3 parallax is clearly wrong ($\mu^1$~Sco = HD 151890).}
Table~\ref{tab:sample} also shows the projected rotational velocity (\vsini), which we took as the median of values in the SIMBAD database.
For some stars without \vsini\ values in SIMBAD, we took measurements directly from the literature \citep{Brown++1997,Wolff++2007,Zorec+Royer2012,Aerts++2014}.

\begin{figure*}
\centering
    \includegraphics[width=1.0\linewidth]{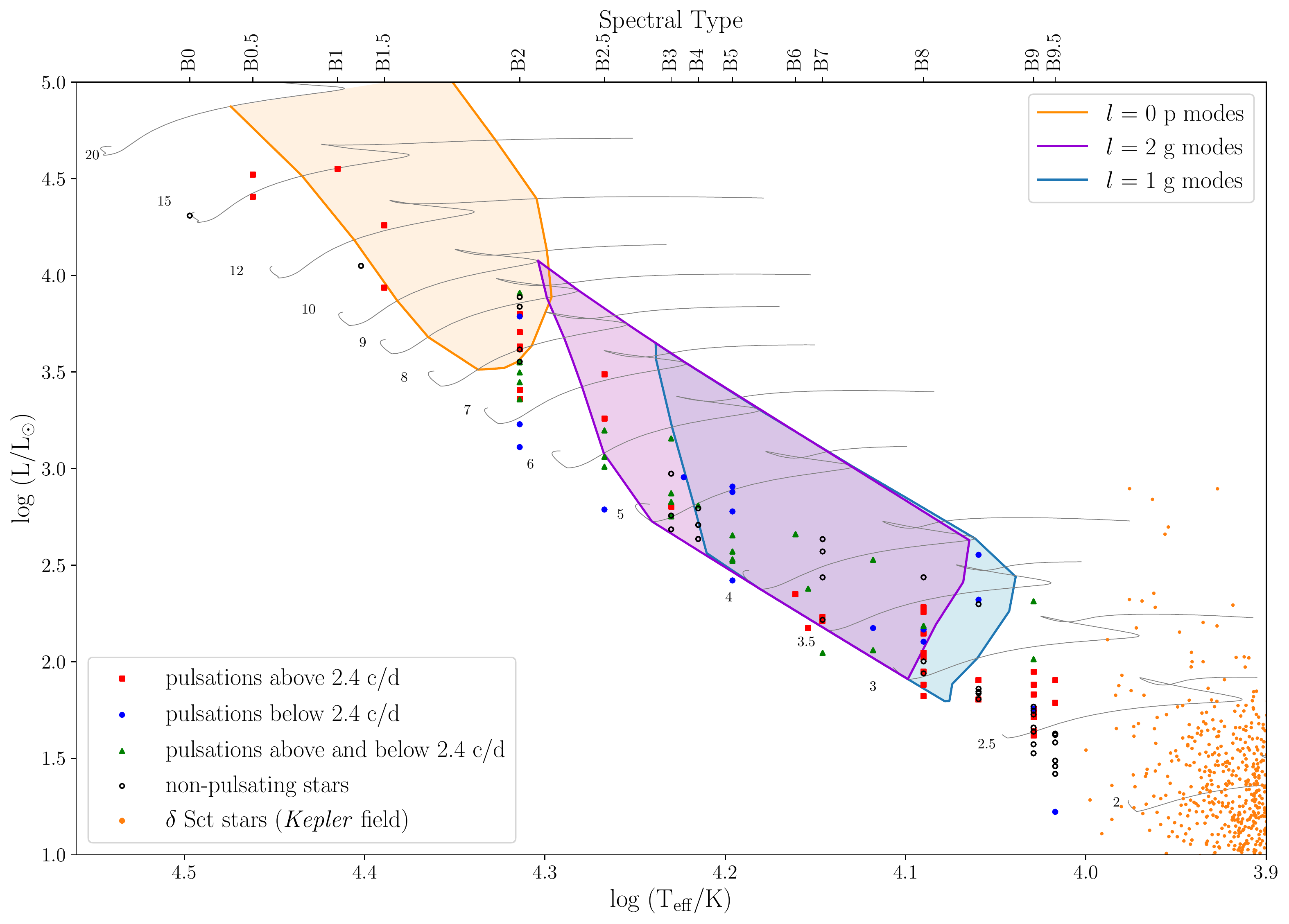}
\caption{H--R diagram showing the variable stars in our Sco--Cen sample. Theoretical instability strips for $l=0$ p~modes and $l=1$ and~2 g~modes are shown in red, blue and purple (see Sec.~\ref{sec:stellar-properties}).  The small orange points show \dsct\ stars observed by the \kepler\ Mission \citep{Murphy2019}. 
}
\label{fig:masterh-r-diagram}
\end{figure*}

The luminosities and effective temperatures of our sample are plotted in a Hertzsprung--Russell diagram in Fig.~\ref{fig:masterh-r-diagram}.  Discussion of some individual stars is given in Sec.~\ref{sec:individual}.  We note that binarity is common among B stars, and so it is likely that properties for some stars are affected by binary companions.  \firstnew{Indeed, \citet{Rizzuto++2013} found that fewer than a quarter of B stars in Sco--Cen are single.} 


The evolutionary tracks in Fig.~\ref{fig:masterh-r-diagram} were calculated using Modules for Experiments in Stellar Astrophysics (MESA version 7184; \citealt{Paxton++2011,Paxton++2013,Paxton++2015}) without rotation, for the chemical composition $(X,Z)=(0.72,0.014)$ and with OPAL opacity tables. For the convective-core boundary we adopted the Schwarzschild criterion and diffusive overshooting with the parameter  $d_{\rm ov} =0.01$.  We estimated the pulsational instability strips for $l=0$ p~modes and $l=1$ and~2 g~modes (red and blue regions in Fig.~\ref{fig:masterh-r-diagram}) by carrying out linear non-adiabatic analysis for selected models using the methods of \citet{Saio+Cox1980} for the nonradial pulsations and of \citet{Saio++1983} for the radial pulsations. 


\section{Analysis of light curves}

The light curves and Fourier amplitude spectra for all stars are shown in the Supplementary on-line material.  \firstnew{For the 23 stars for which we used 30-minute FFIs (see Sec.~\ref{sec:lightcurves}), the light curves are shown connected by a solid line rather than with dots, and the Nyquist frequency of 24\,\cd\ is shown in the amplitude spectrum by a vertical dashed line.}

Five of the stars \new{have previously been classified as eclipsing binaries or ellipsoidal variables}, and these are discussed individually in Sec.~\ref{sec:eclipsing-binaries}.  Other light curves show variations typical of rotation or close binary companions, such as 13~Sco, V863~Cen, V1019~Cen, 52~Hya, KT~Lup, HD~132094 (a heartbeat star with a very short period; see Sec.~\ref{sec:hd-132094}), HD~136482, and HD~143022.  

\begin{figure} 
\centering
    \includegraphics[width=\columnwidth]{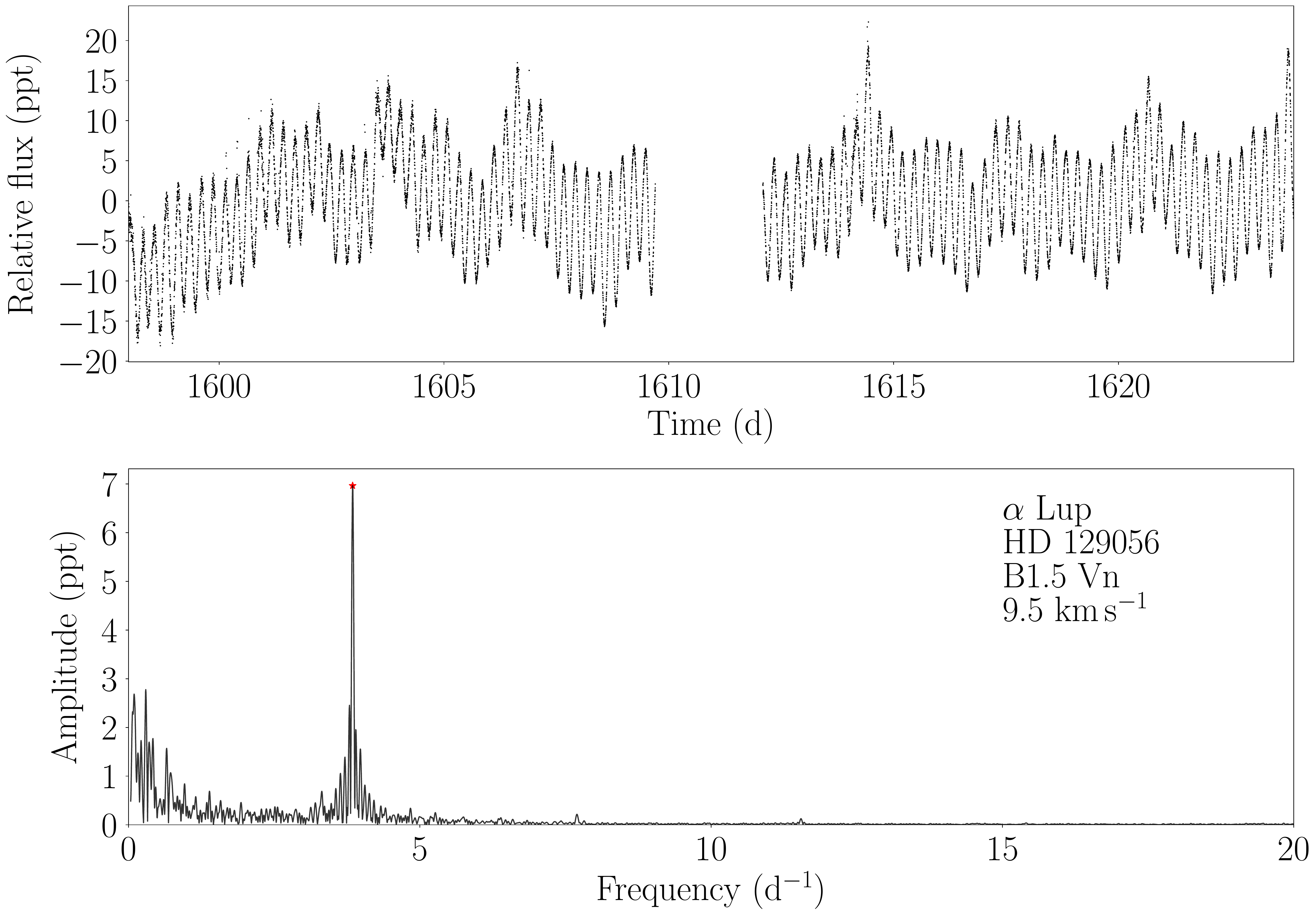}
\caption{TESS light curve and Fourier amplitude spectra of $\alpha$~Lup (HD~129056; B1.5~Vn), which is a previously-known \bcep\ star \citep{Nardetto++2013}.}
\label{fig:example-bcep}
\end{figure}

\begin{figure} 
\centering
    \includegraphics[width=\columnwidth]{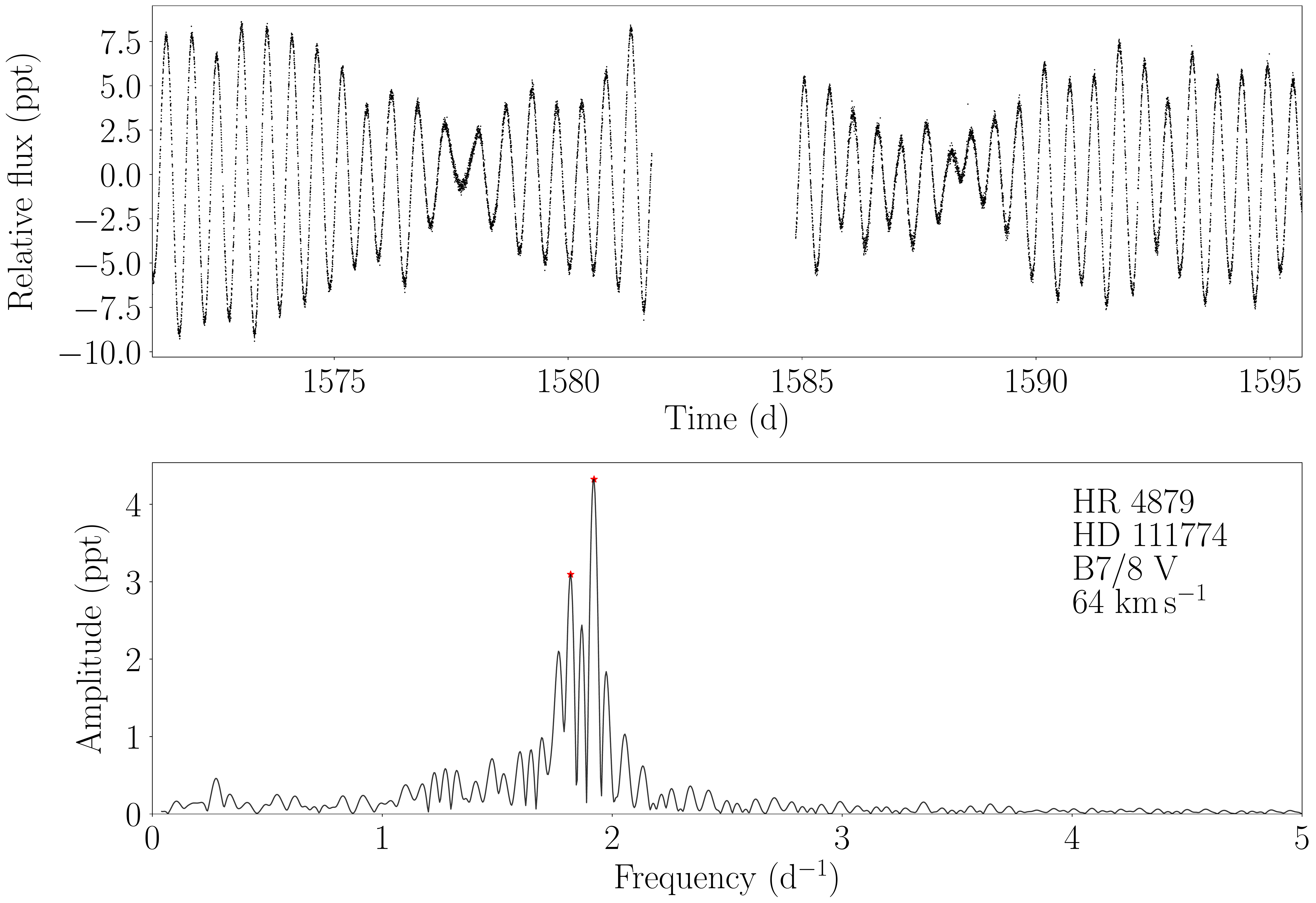}
\caption{TESS light curve and Fourier amplitude spectra of HR~4879 (HD~111774; B7/8~V), an SPB star with two closely-spaced modes.}
\label{fig:example-spb}
\end{figure}

\begin{figure} 
\centering
    \includegraphics[width=\columnwidth]{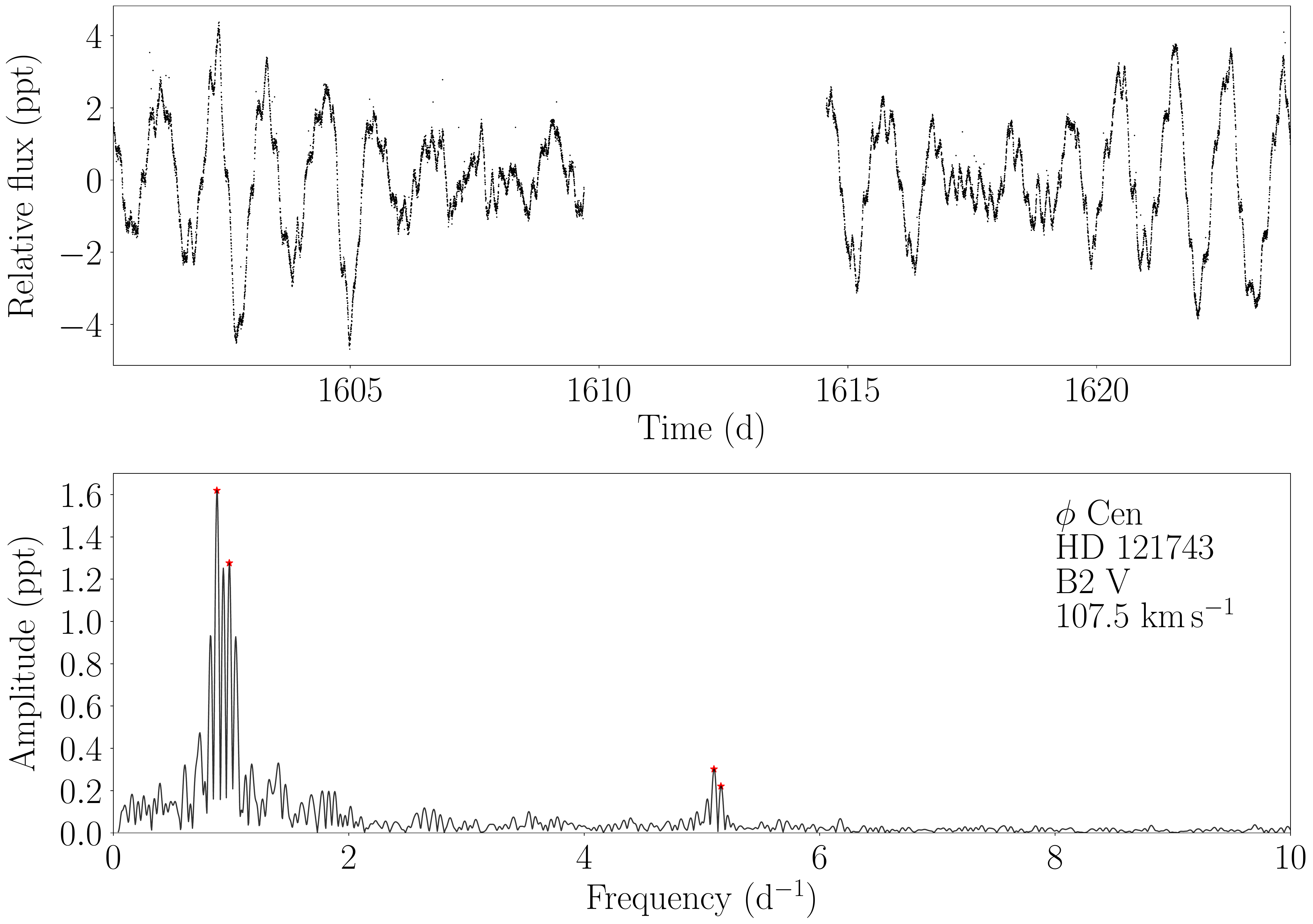}
\caption{TESS light curve and Fourier amplitude spectra of $\phi$~Cen (HD~121743; B2~V), which is a hybrid pulsator (\bcep\ and SPB).}
\label{fig:example-hybrid}
\end{figure}

\begin{figure}
\centering
    \includegraphics[width=\columnwidth]{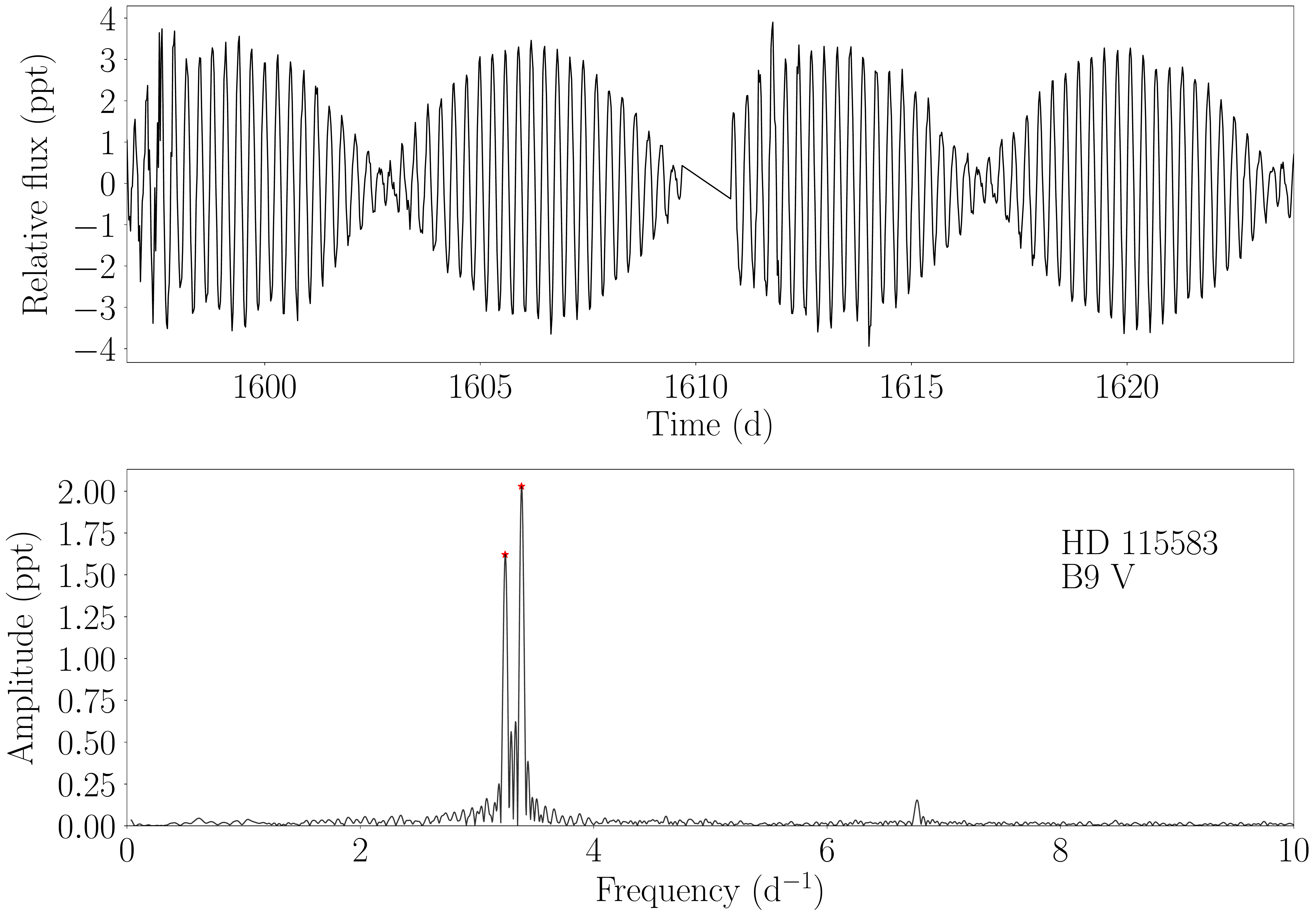}
\caption{TESS light curve and Fourier amplitude spectra of HD~115583 (B9~V), which shows clear pulsations with two closely-spaced modes but is cooler that the standard SPB instability strip.}
\label{fig:example-cool-spb}
\end{figure}

Many stars show clear pulsations and we include four examples in Figs.~\ref{fig:example-bcep}--\ref{fig:example-cool-spb}. 
\firstnew{In each figure we give the name, HD number, spectral type and (where available) the value of \vsini.}
Figure~\ref{fig:example-bcep} shows the previously-known \bcep\ star $\alpha$~Lup (HD~129056; B1.5~Vn; \citealt{Nardetto++2013}).
Figure~\ref{fig:example-spb} shows HR~4879 (HD~111774; B7/8~V), which we find to be an SPB star with two closely-spaced modes.
Figure~\ref{fig:example-hybrid} shows $\phi$~Cen (HD~121743; B2~V), which is seen to be a hybrid pulsator (\bcep\ and SPB).
Finally, Fig.~\ref{fig:example-cool-spb} shows HD~115583 (B9~V), which shows clear pulsations with two closely-spaced modes but is cooler than the standard SPB instability strip.
We note that fine structure in the oscillation spectra of SPB stars that has been seen with long runs with CoRoT and \kepler\ \citep[e.g.][]{Degroote++2010,Papics++2017} is not fully resolvable with only two sectors (54\,d) of TESS data.



In the H--R diagram (Fig.~\ref{fig:masterh-r-diagram}), we have indicated the 81 stars that show pulsations (68\% of the sample).  The theoretical instability strips for \bcep{} and SPB stars are overlaid.
For comparison, we also show the sample of \dsct\ stars observed by \kepler\ (small orange points; \citealt{Murphy2019}).  

To assist the comparison with the theoretical instability strips for \bcep{} and SPB stars, we have indicated whether pulsations fall above or below a somewhat arbitrary threshold of 2.4\,\cd.  We did not attempt a detailed frequency analysis.  Instead, for each star that shows probable pulsations, we have listed up to four frequencies in Table~\ref{tab:sample}.  In particular, we list the two strongest pulsation peaks both above and below the 2.4-\cd\ threshold (and excluding harmonics, \new{which we have labelled with `H' in the amplitude spectra in the on-line supplementary material}).  
We note that it can be difficult to distinguish between rotation and low-frequency pulsation in B stars \citep[e.g.,][]{Briquet++2007,Lee2021}, and it is possible that some stars listed as pulsating are actually rotational variables, and vice versa.  \new{In the figures, we have used `R' to label peaks that we attributed to rotation rather than pulsation, with the caveat that some of these may need to be revised. We have also added the word `binary' on plots where the low-frequency peaks in the amplitude spectrum are due to eclipses or ellipsoidal variations.}

\firstnew{The uncertainties on the measured frequencies depend on the duration of the data set ($T$) and the signal-to-noise ratio (SNR) of the peak in the amplitude spectrum.  Specifically, we expect the frequency uncertainty to be approximately $0.44\,{\rm SNR}^{-1}\,T^{-1}$    
\citep{Montgomery+ODonoghue1999, Kjeldsen+Bedding2012}.  For light curves spanning a single \tess\ sector ($T=27$\,d, which applies to most stars in Table~\ref{tab:sample}) and peaks with SNR in the range 50 down to 5, the frequency uncertainty is therefore about 0.0003--0.003\,\cd.  These uncertainties are much smaller than the symbol sizes on the plots in this paper.}

\section{Discussion}
\label{sec:discussion}

\begin{figure*}
\centering
     \includegraphics[width=1.85\columnwidth]{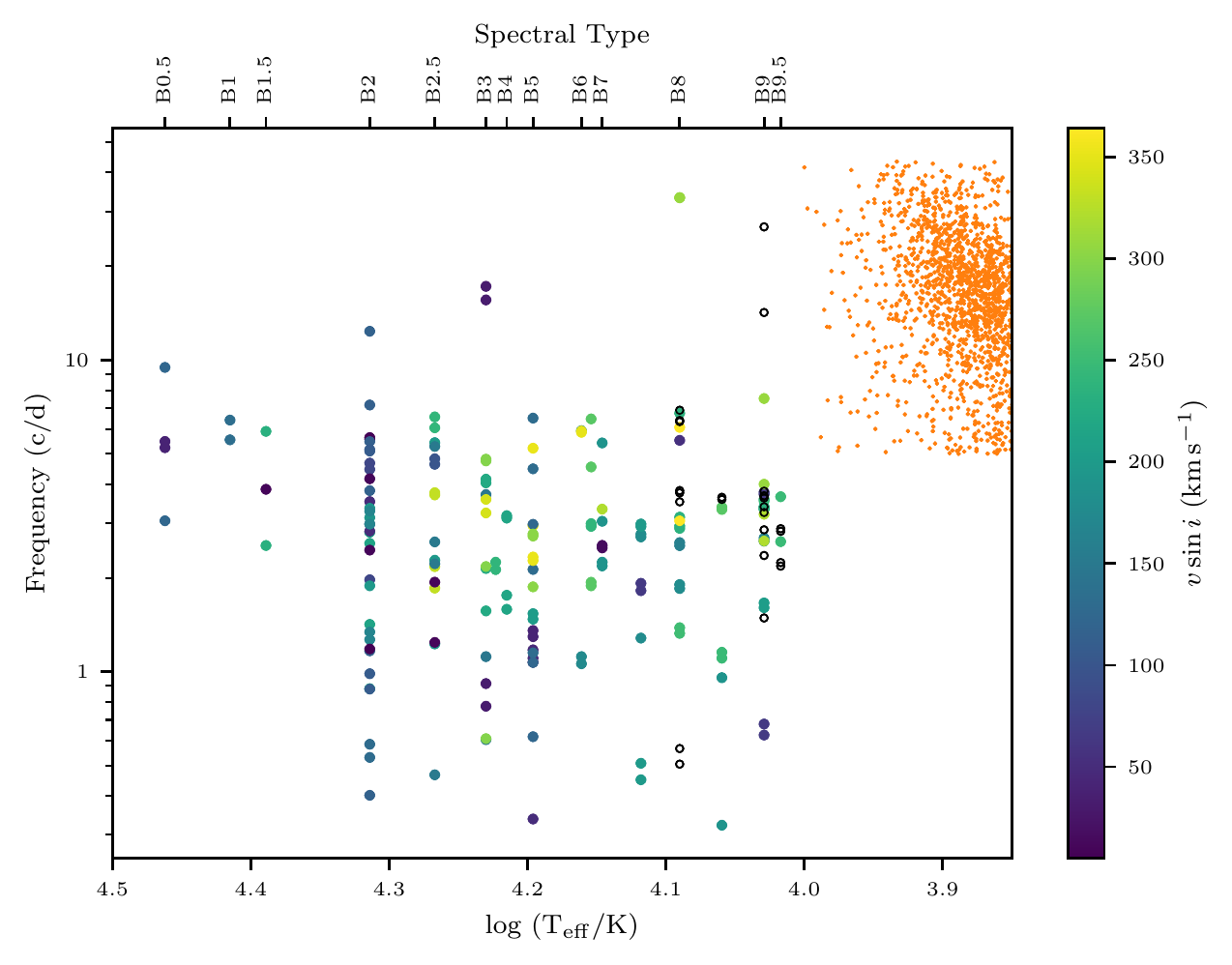}
\caption{Pulsation frequencies of B~stars in Sco--Cen from \tess\ observations, using the values listed in Table~\ref{tab:sample}. Symbols are colour-coded with \vsini, and left as open circles when this is not available.  The small orange points show \dsct\ stars observed by \kepler\ \citep{Murphy2019}. }
\label{fig:samplefrequencyvsini}
\end{figure*}

We see from Fig.~\ref{fig:masterh-r-diagram} that pulsations occur across the full range of effective temperatures \citep[see also][]{Pedersen++2019,Balona+Ozuyar2020}.  They do not seem to correspond exactly with the theoretical instability strips, although we note that (i)~some effective temperatures and luminosities could be in error, and (ii)~the boundaries of the theoretical instability regions are sensitive to the heavy element abundance ($Z$), the opacity tables and rotation
\citep[e.g.,][]{Pamyatnykh1999,Townsend2005,Paxton++2015,Moravveji2016,Saio++2017}.
In Fig.~\ref{fig:samplefrequencyvsini} we show the pulsation frequencies listed in Table~\ref{tab:sample} as a function of effective temperature, with the symbols colour-coded by the \vsini\ of the star. \firstnew{We again see that pulsations are present across the entire range of effective temperatures, with an overall tendency for lower-frequency pulsations to occur in cooler stars. We also see that the standard threshold of about 2--3\,\cd\ is not by itself a reliable way to distinguish \bcep\ stars (p modes) from SPB stars (g modes).   }


\begin{figure*}
\centering
    \includegraphics[width=0.9\linewidth]{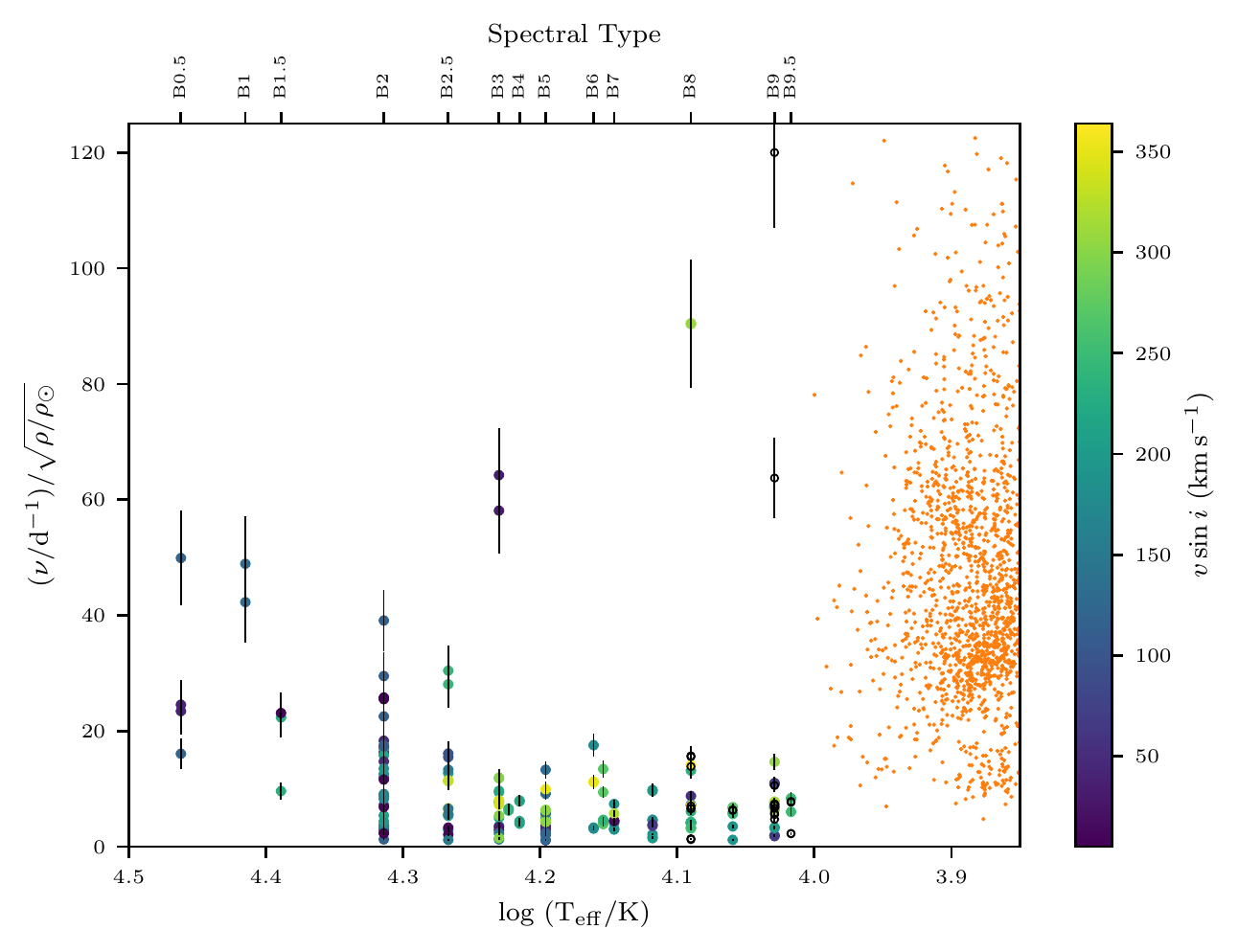}
\caption{Scaled pulsation frequencies as a function of stellar effective temperature.
Similar to Fig.~\ref{fig:samplefrequencyvsini} except that the frequencies have been divided by the square root of the mean stellar density (see text). Densities for the \kepler\ \dsct\ stars (small orange points) were calculated using masses and radii from \citet{Murphy2019}.   
}
\label{fig:logQvslogTvsini}
\end{figure*}

To decide whether the pulsations in a given star are p- or g-modes, it is very helpful to know the mean density of the star.  This is because the frequency of a particular p mode, such as the radial fundamental mode, scales from one star to another as the square root of the mean stellar density. Indeed, when plotting theoretical models it is common to scale the frequencies by dividing by $\sqrt{\rho}$ \citep[e.g.][]{Aerts++2010-book}.  We show this for the observed pulsation frequencies in Figure~\ref{fig:logQvslogTvsini}, which is the same as Fig.~\ref{fig:samplefrequencyvsini} except that the frequencies have been divided by the square root of the stellar density.  
We made these rough density estimates for the stars in our sample as follows.  

\begin{figure}
\centering
    \includegraphics[width=\columnwidth]{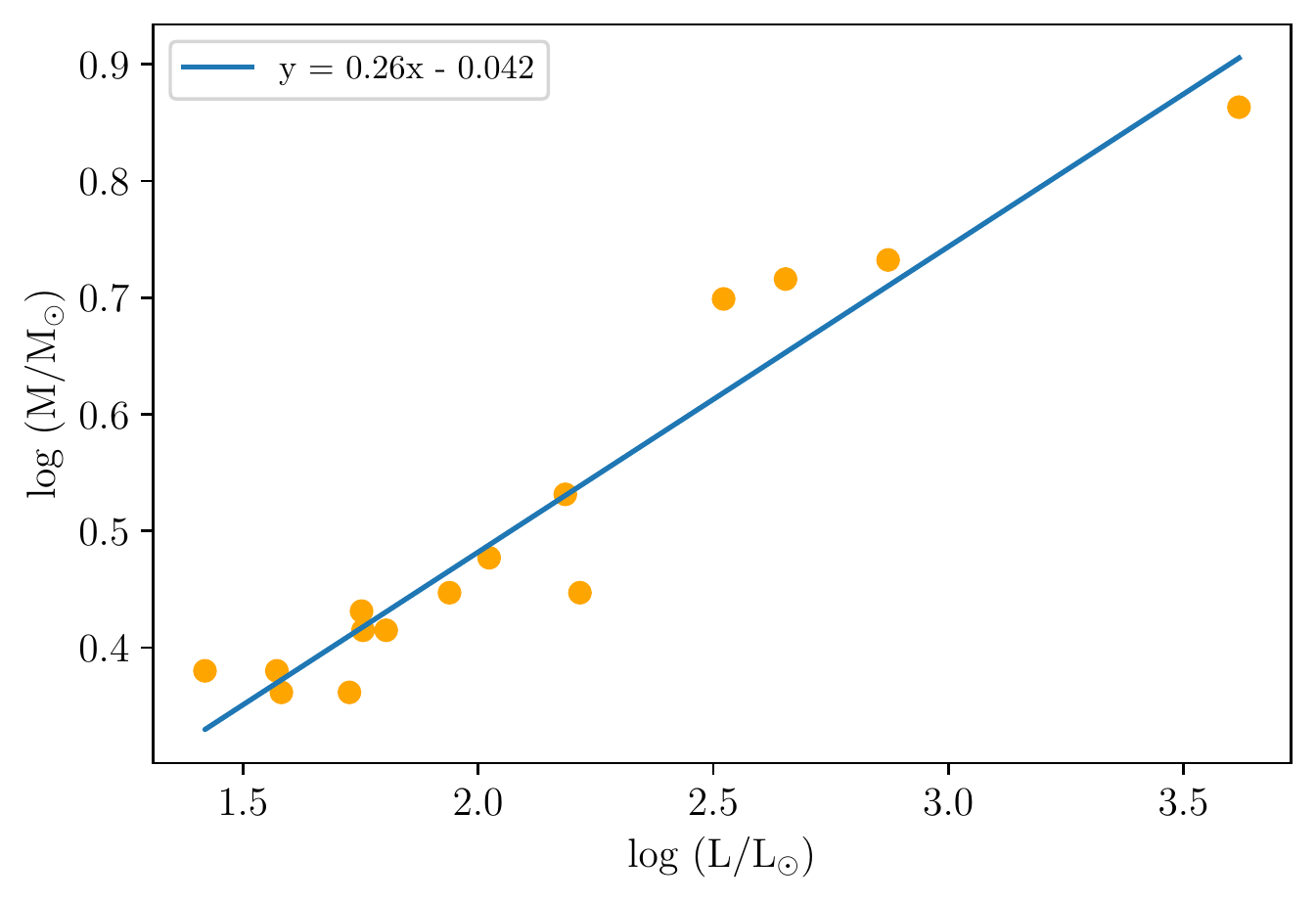}
\caption{Mass--luminosity relationship used to estimate masses of stars in our sample. The masses for these 15 stars are taken from \citet{Jang++2015}.
}
\label{fig:M--L}
\end{figure}

For each star, we estimated the radius using the Stefan--Boltzmann law ($ L \propto R^2 \Teff^4$), using the values in Table~\ref{tab:sample}.  We estimated masses from luminosities using a mass--luminosity relation 
\citep{Eker++2018}, which we calibrated for Sco--Cen using published masses for 15 stars in our sample \citep{Jang++2015}, as shown in Fig.~\ref{fig:M--L}.  We found the relation to be
\begin{equation}
    \log{M/\rm M_{\odot}} = (0.26 \pm 0.02) \log{L/\rm L_{\odot}} - (0.042 \pm 0.045),
    \label{eq:M-Lrelationship2}
\end{equation}
which we used to estimate the masses for our sample.

The relationship between pulsation periods and mean density is usually written as \citep[e.g.,][Sec.~5.2]{Catelan+Smith2015}:
\begin{equation}
    P = Q \left(\frac{\rho}{\rho_\odot}\right)^{-0.5}. \label{eq:Q}
\end{equation}
The quantity $Q$, called the pulsation constant, depends on the pulsation mode.  
For example, $Q$ is about 0.033\,d for the fundamental radial mode in \dsct\ stars \citep[e.g.,][]{Breger2000}.
It follows from Eq.~\ref{eq:Q} that the vertical axis in Fig.~\ref{fig:logQvslogTvsini} shows $1/Q$, where $Q$ is the pulsation constant (in days).  

\citet{Stankov+Handler2005} suggested that $Q$ for \bcep\ stars peaks at 0.033\,d (the same value as \dsct\ stars), which corresponds to $1/Q=30.0\,\cd$.  To verify this, we used the grid of stellar models calculated by \citet{Stello++2009} using the Aarhus stellar evolution code \citep[ASTEC;][]{C-D2008}, and confirmed that the fundamental radial mode satisfies $1/Q=30.2\,\cd$ across a range of masses and ages.
In principle, all pulsations falling below this value cannot be p modes.  However, we must remember that there are uncertainties in masses, both from uncertainties in luminosities and from scatter in the mass-luminosity relation (Fig.~\ref{fig:M--L}).  An error of 0.2 in $\log M$, which is an extreme case, corresponds to a change of 0.1 in $\log(1/Q)$, which is a change of up to $\sim$25\% in $1/Q$.  


We see in Fig.~\ref{fig:logQvslogTvsini} that \new{with three exceptions}, all stars with spectral type cooler than B2 show pulsations that fall below the p-mode limit, which implies they must be g modes (or perhaps Rossby modes).  \new{The first exception is $\xi^2$~Cen (HD~113791; B3) which, as discussed in Sec.~\ref{sec:xi2-cen}, has an unusually regular p-mode spectrum.  The second is HR~4874 (HD~111597; B9\,V), whose amplitude spectrum (Supplementary on-line material) has many peaks above 10\,\cd\ and this star appears to be a \dsct\ pulsator (SIMBAD lists other references that give a spectral type of A0).}  The third exception is $\mu$~Lup (HD~135734; B8\,Ve), which is a Be star \citep{Levenhagen+Leister2006, Arcos++2018} and so its anomalously high-frequency variations (33\,\cd) presumably come from oscillations in the circumstellar disk. 

Our statement that pulsations with $1/Q$ below about 30\,\cd\ cannot be p~modes is particularly important for the stars that lie between the accepted SPB and \dsct\ instability strips (spectral types B9 and B9.5).  As mentioned in the Introduction, there have been several reports of pulsations in B stars that are cooler than the standard range for SPB stars.  \citet{Salmon++2014} suggested the examples found by \citet{Mowlavi2013} in the young open cluster NGC~3766 are actually fast-rotating SPB pulsators, and this was endorsed by \citet{Saio++2017}.  The idea, which was shown theoretically by \citet{Townsend2005}, is that the instability strip for prograde sectoral modes in SPB stars is shifted towards lower luminosities and effective temperatures by rapid rotation.  Looking at the \vsini\ values for our sample (colour-coded in Fig.~\ref{fig:logQvslogTvsini}), we indeed see that \firstnew{the majority of} stars with spectral types B9 and B9.5 \firstnew{for which \vsini\ measurements are available} are indeed rotating rapidly (\vsini\ is in the range 150 to 320\,\kms\ \firstnew{for 16 of the 17 stars that have measurements}, \new{and 8 of the 9 among these in which we detected pulsations.}).


\subsection{Period--luminosity relations for fast rotators}
\label{sec:p-l-relations}

\citet{Mowlavi2016} reported that the fast rotators in NGC~3766 followed two distinct period--luminosity (P--L) relations. \citet{Saio++2017} suggested that these two relations corresponded to prograde sectoral g modes that are dipole ($l=m=1$) and quadrupole ($l=m=2$), respectively\footnote{we adopt the convention that prograde modes have $m>0$}. 
%
%
%
It is known that g-mode oscillations observed in a rapidly rotating stars are predominantly prograde sectoral modes \citep{Van-Reeth++2016,Ouazzani++2017,Papics++2017,Li++2019-catalogue}. In such cases, the pulsation frequencies of g modes in the observer's frame are given as $\nu_{\rm obs} = \nu_{\rm corot} + mf_{\rm rot} \approx mf_{\rm rot}$. 
The latter relation is obtained assuming the frequencies of g modes in the co-rotating frame $\nu_{\rm corot}$ to be much smaller than the rotation frequency $f_{\rm rot}$; i.e., $\nu_{\rm corot} \ll  mf_{\rm rot}$. In other words, these g modes in rapidly rotating stars should have periods similar to $P_{\rm rot}/m$.

\begin{figure*}
\centering
    \includegraphics[width=2\columnwidth]{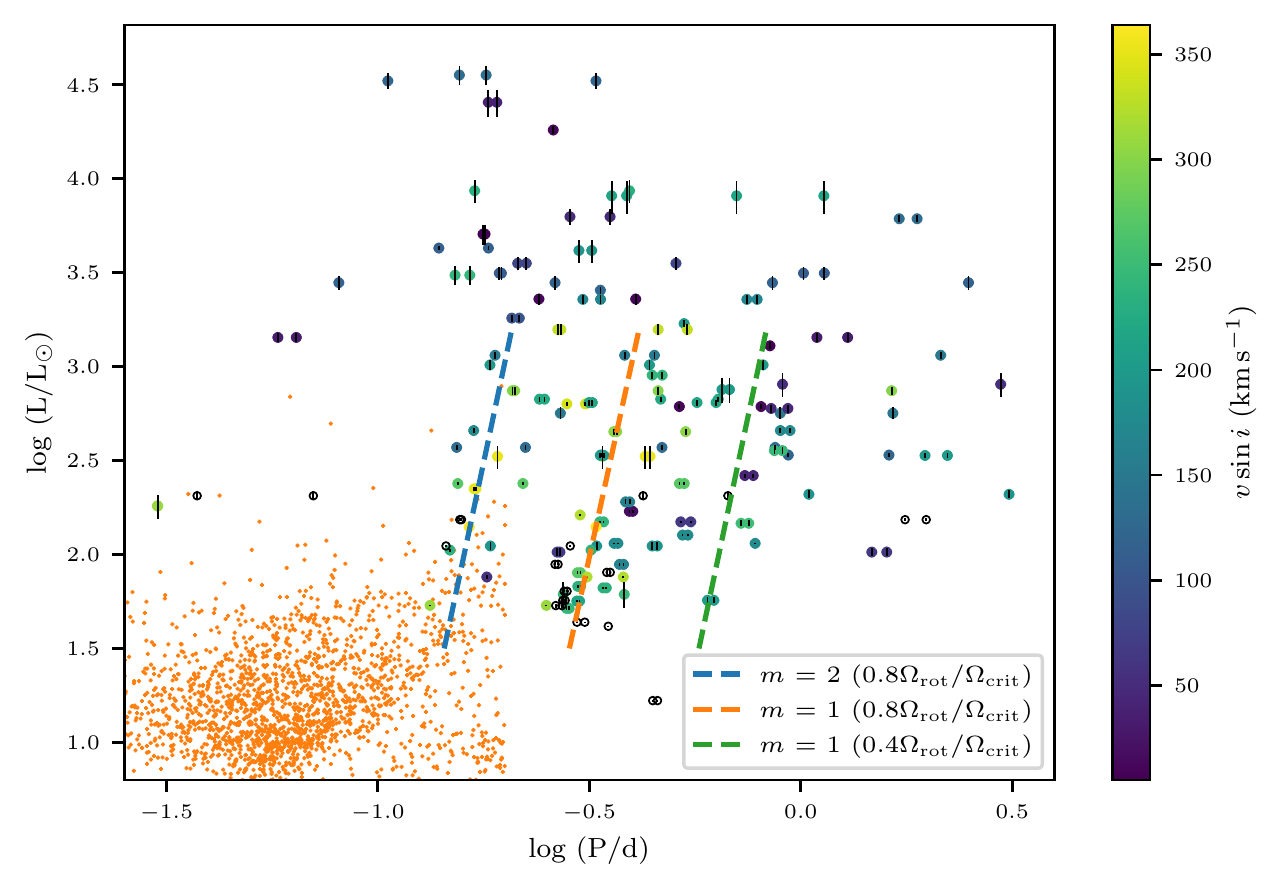}
\caption{Period--luminosity diagram for the stars in our Sco--Cen sample (colour-coded with effective temperature), as well as \dsct\ stars observed by \kepler\ (orange points; \citealt{Murphy2019}).  Dashed lines show theoretical P--L relations for sectoral g~ modes for rotation rates of 0.4 and 0.8 times the critical value (see Sec.~\ref{sec:p-l-relations}).
}
\label{fig:p-l-diagram}
\end{figure*}

Figure~\ref{fig:masterh-r-diagram} shows that most SPB stars in the Sco--Cen association lie close to ZAMS, reflecting their very young ages.  They are therefore expected to be rotating close to the critical rates. Assuming that they rotate at rates proportional to the critical rate of the ZAMS model corresponding to that luminosity,  we can connect the rotation rate (and hence pulsation period $P_{\rm obs} \approx P_{\rm rot}/m$) with the luminosity as
\begin{multline}
    \log(P_{\rm obs}/ {\rm d}) \approx - \log(\Omega_{\rm rot}/\Omega_{\rm crit}) - 0.60 + \\ 0.094(\log(L/\Lsun) - 1.91)  - \log(m).
    \label{eq:PL}
\end{multline}
Here, $\Omega_{\rm rot}/\Omega_{\rm crit}$ expresses the rotation frequency as a fraction of its critical value.
Equation~\ref{eq:PL} represents a period--luminosity relation for given values of $\Omega_{\rm rot}/\Omega_{\rm crit}$ and $m$.  
\firstnew{Equation~\ref{eq:PL} was obtained by a linear interpolation using $\log L$ and $\log R$ of ZAMS models with masses of $3\Msun$ and $6\Msun$ (see also \citealt{Saio++2017}), and can be applied for rotation rates greater than about 0.4 of the critical value. 
We verified that the equation is valid to within 0.01 for $1.6 \le \log{L/\Lsun} \le 3.0$, which covers the range of g-mode pulsators.  }

The P--L relations for $\Omega_{\rm rot}/\Omega_{\rm crit}=0.4$ and 0.8 are shown in Fig.~\ref{fig:p-l-diagram} by dashed lines.  Many of the rapidly rotating SPB stars fall \firstnew{in the region expected from} these relations, \firstnew{consistent with} our prediction. 
Furthermore, this figure shows that variable stars cooler and fainter than the standard SPB instability boundary ($1.5 \le \log(L/\Lsun) \le 2.0$; see Fig.~\ref{fig:masterh-r-diagram}) fall approximately on the $m=1$ P--L relations, indicating they are \new{likely to be} rapidly rotating SPB stars (g-mode pulsators). \new{It would clearly be valuable to obtain \vsini\ measurements for the stars that lack them.}

The PL relations for rapidly rotating SPB stars in the Sco-Cen association we found are the same type as the relations for rapidly rotating SPB stars in the young open cluster NGC 3766 found by \citet{Mowlavi2016} and modeled by \citet{Saio++2017}. The presence of such P--L relations in the young cluster and the OB association indicates many newly born B-type (single) stars rotate nearly critically.   

\section{Notes on individual stars} \label{sec:individual}

\subsection{\texorpdfstring{$\alpha$}{alpha}~Cru: \firstnew{confirmation of} \texorpdfstring{$\beta$}{beta}~Cephei \firstnew{pulsations}}
\label{sec:alpha-cru}

The brightest star in the Southern Cross is a multiple system whose two most luminous components are $\alpha^1$~Cru (HD~108248; B0.5\,V; $V = 1.28$) and $\alpha^2$~Cru (HD~108249; B1\,V; $V = 1.58$).  They are separated by only 4\arcsec\ and so the \tess\ light curve includes both (\tess\ pixels are 21\arcsec).  As described in Sec.~\ref{sec:lightcurves}, the extreme brightness of this target required us to use a pixel-level analysis to extract the light curve. 

\begin{figure}
\centering
    \includegraphics[width=\columnwidth]{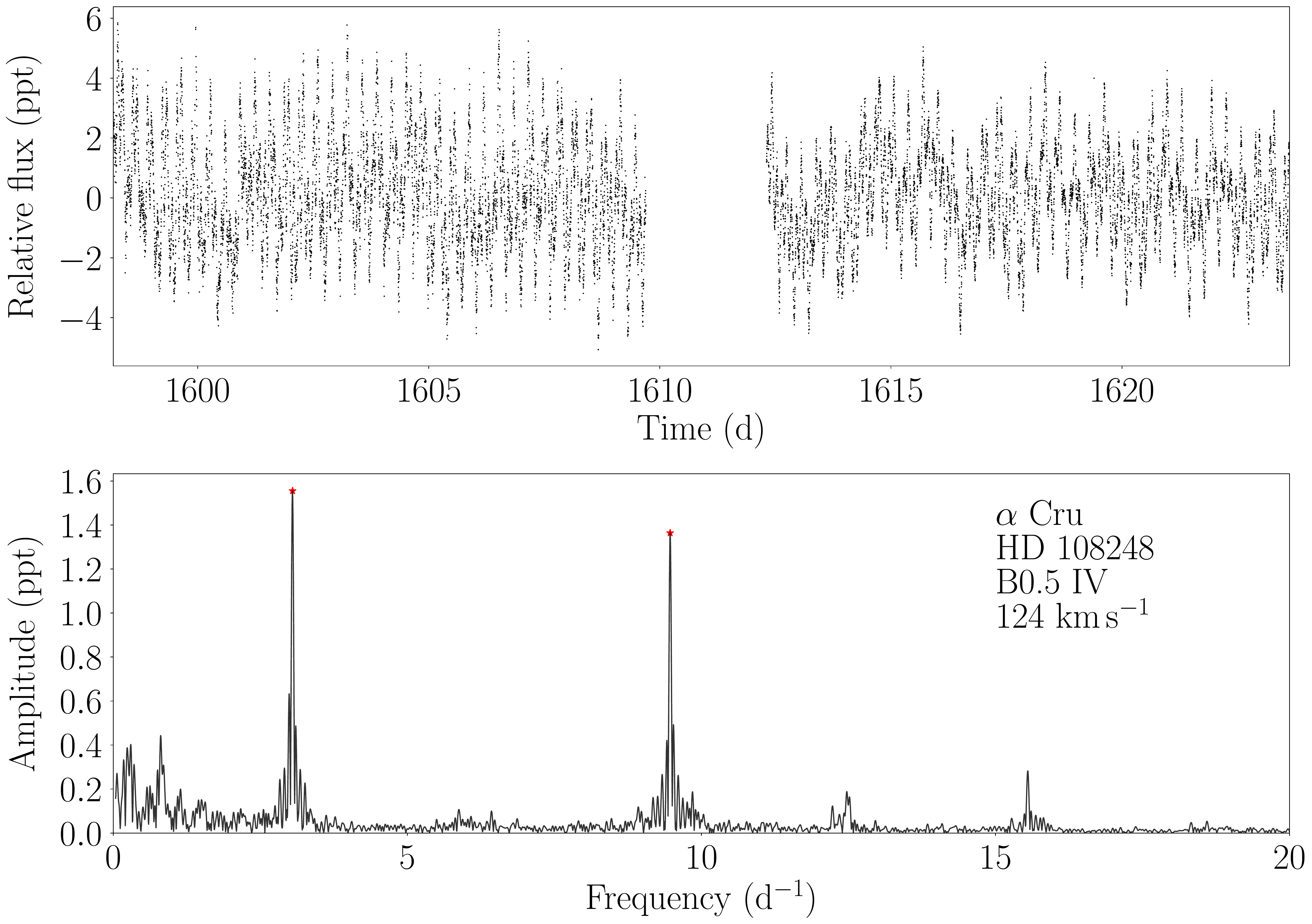}
\caption{\tess\ light curve and amplitude spectrum of the \bcep\ star $\alpha$~Cru (see Sec.~\ref{sec:alpha-cru}).
}
\label{fig:alpha-cru}
\end{figure}

Our \tess\ light curve (Fig.~\ref{fig:alpha-cru}) clearly shows low-amplitude pulsations, making it the second-brightest \bcep\ star in the sky (after $\beta$~Cen). The \bcep\ variability of $\alpha$~Cru from the \tess\ data was also noted in Table~A.1 of \citet{Bowman++2022}, \firstnew{although no frequencies were given}.  The amplitude spectrum in Fig.~\ref{fig:alpha-cru} shows modes at frequencies  3.048\,\cd\ (amplitude 0.25\,ppt) and 9.468\,\cd\ (0.21\,ppt), which were previously reported by \citet{Kolaczek-Szymanski2020}, based on photometry from the SMEI and BRITE spacecraft.
There are additional peaks at combination frequencies of the two strongest peaks, the strongest being at 15.548\,\cd\ (0.05\,ppt). 
\new{There appears to be a regularity and fine structure in the spectrum, and this star is clearly worthy of further study.}
It is not possible to tell definitively whether the pulsations occur in $\alpha^1$~Cru or $\alpha^2$~Cru (or both), although the presence of combination frequencies shows that the two strongest modes must both occur in the same star.  For the figures in this paper, we used the parameters of $\alpha^1$~Cru.  


\subsection{\texorpdfstring{$\beta$}{beta}~Cen: a multi-mode \texorpdfstring{$\beta$}{beta}~Cephei star}
\label{sec:beta-cen}

\begin{figure}
\centering
    \includegraphics[width=\columnwidth]{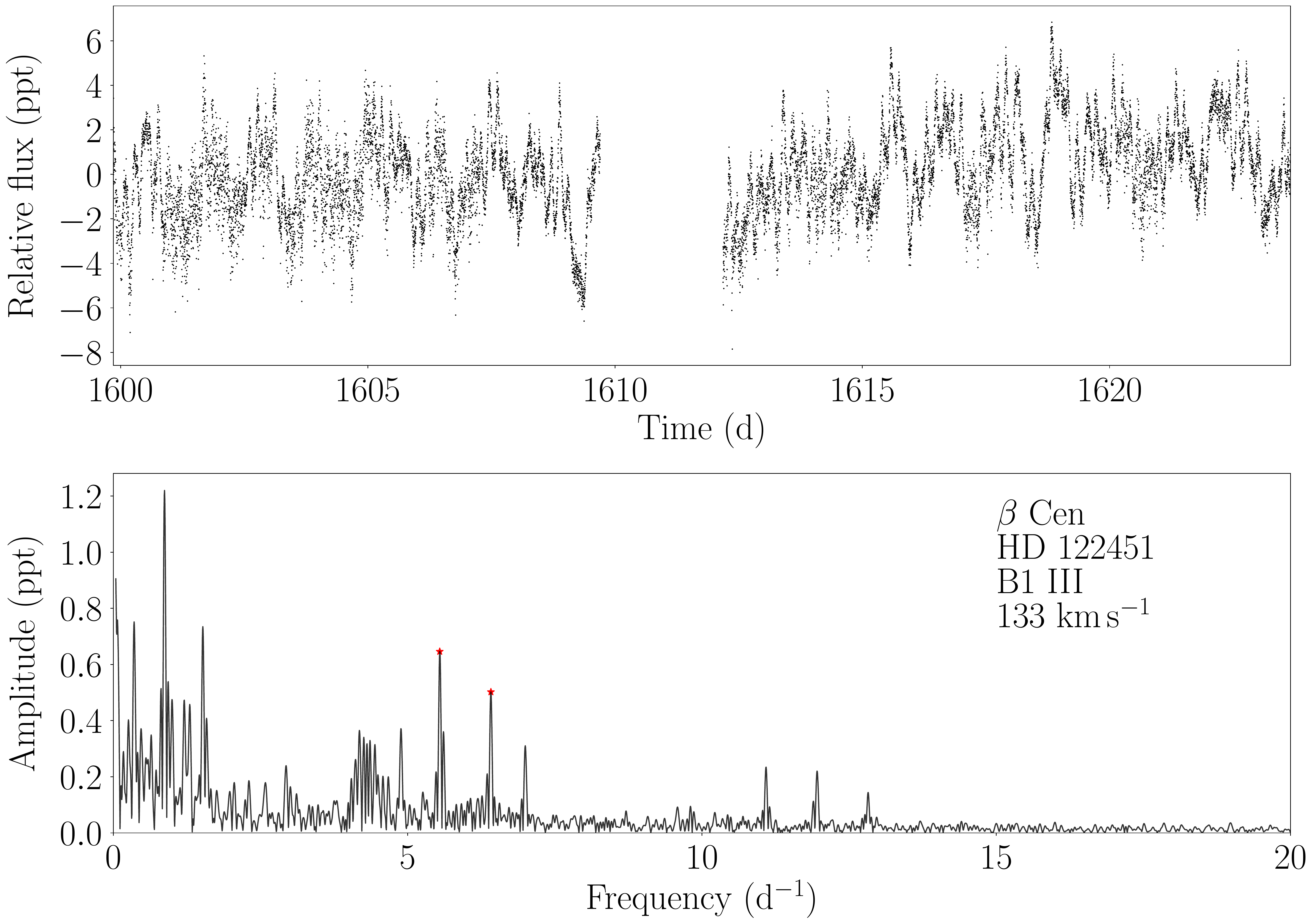}
\caption{\tess\ light curve and amplitude spectrum of the \bcep\ star $\beta$~Cen (see Sec.~\ref{sec:beta-cen}).
}
\label{fig:beta-cen}
\end{figure}
$\beta$~Cen (HD~122451; B1\,III) is known to show multi-periodic pulsations, based on photometry spanning 146\,d from the BRITE-Constellation \citep{Pigulski2016++}. It is an interferometric and double-lined spectroscopic binary with two nearly-equal components \citep{Robertson++1999,Ausseloos++2006}, and our luminosity is calculated from an absolute magnitude for the primary of $M_V=-4.03$ \citep{Pigulski2016++}.
Our \tess\ light curve, extracted from pixel-level data (Sec.~\ref{sec:lightcurves}), confirms a rich spectrum of pulsation modes \firstnew{(see Fig.~\ref{fig:beta-cen})}, including some that were not detected with BRITE (the BRITE spacecraft are in low-Earth orbit, which results in an effective Nyquist frequency of 7.17\,\cd; see \citealt{Pigulski2016++}).  
The TESS amplitude spectrum \firstnew{shows additional peaks around 12\,\cd\ that are combination frequencies of the peaks around 5--6\,\cd.}


    


\subsection{\texorpdfstring{$\xi^2$}{xi2}~Cen: a hybrid pulsator with regular spacings}
\label{sec:xi2-cen}

\begin{figure}
\centering
    \includegraphics[width=\columnwidth]{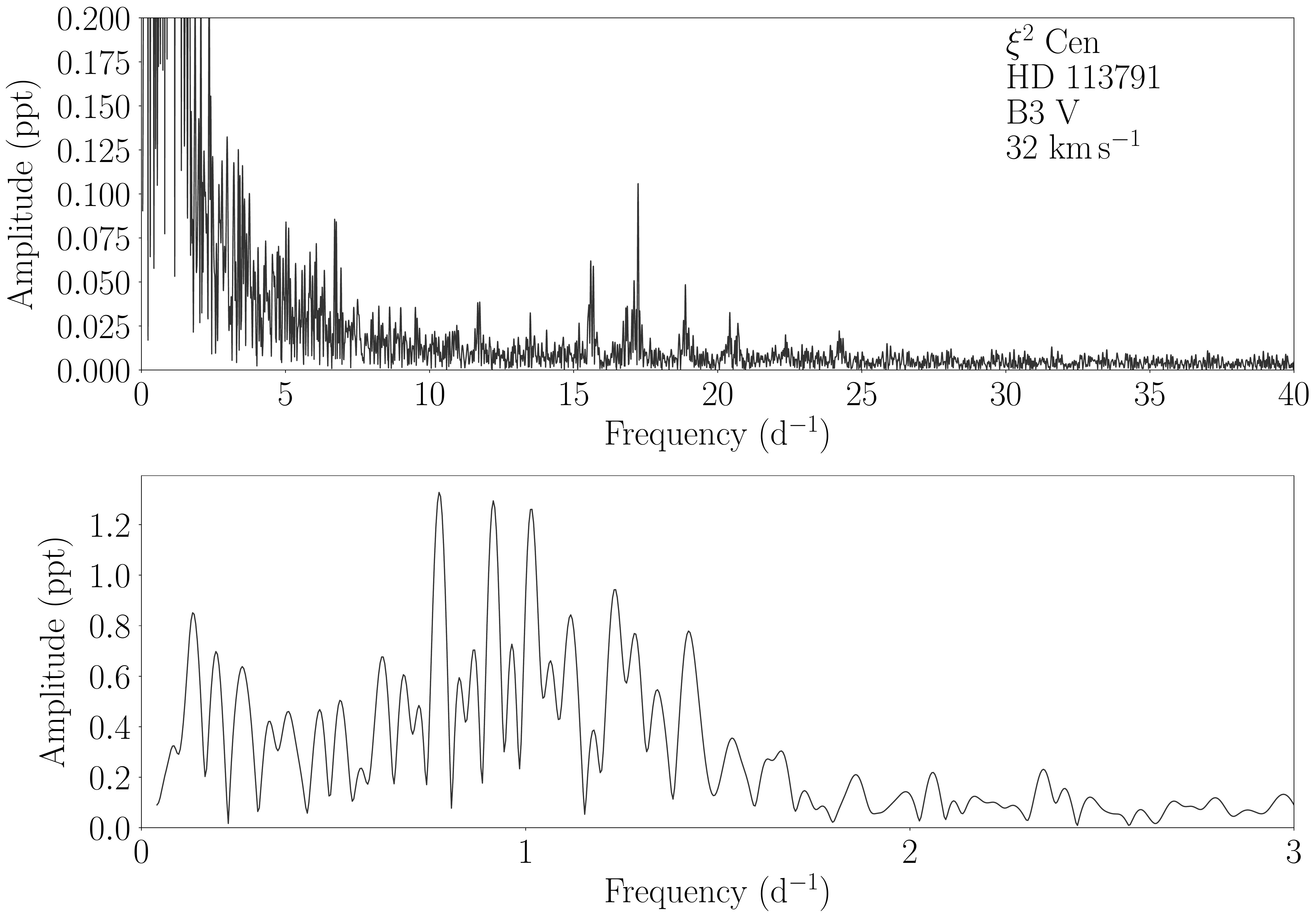}
    \includegraphics[width=0.8\columnwidth]{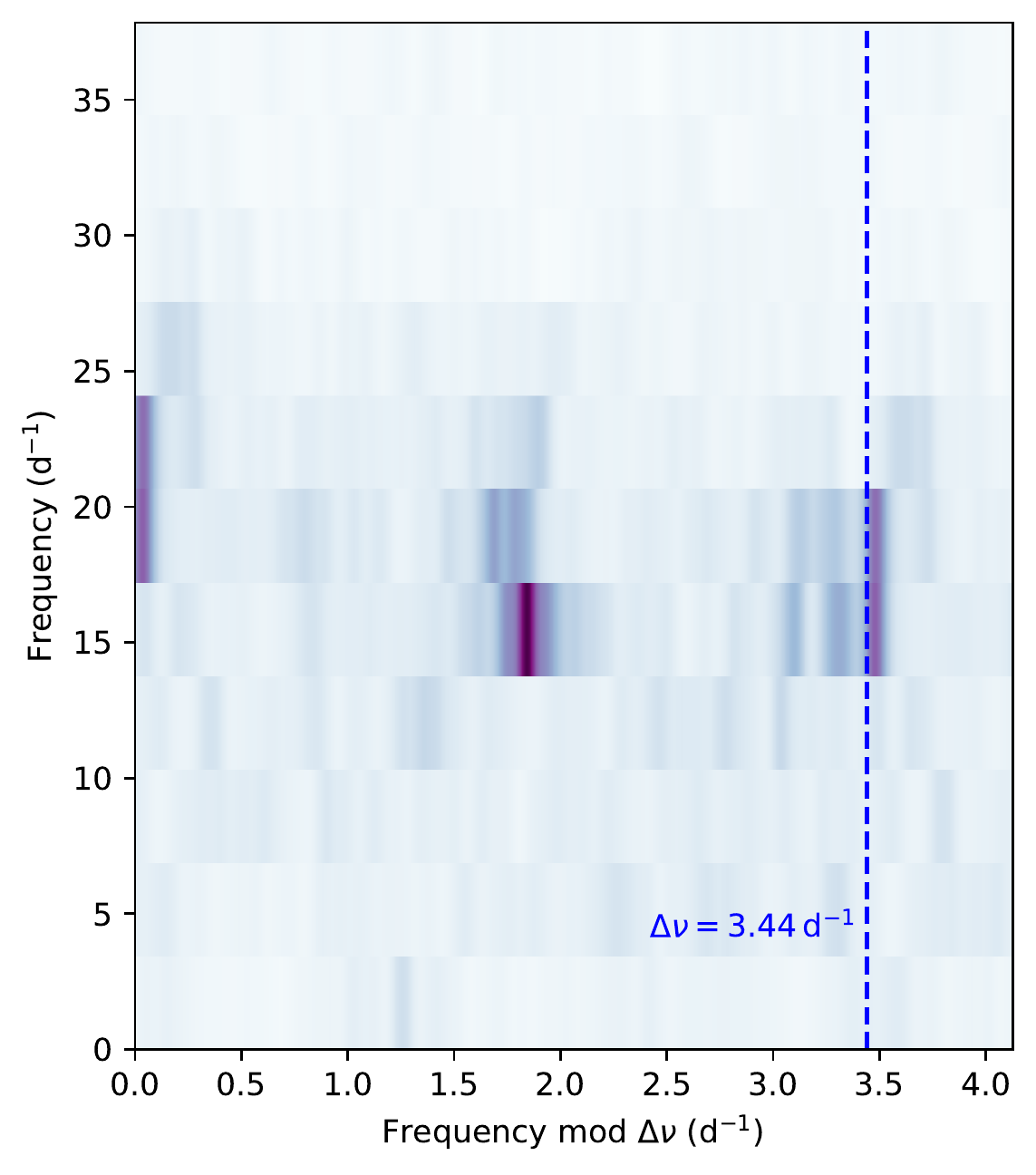}
\caption{\tess\ amplitude spectrum of the hybrid pulsator $\xi^2$~Cen from \tess\ Sectors 11, 37 and 38, showing the regions with p~modes (top panel) and g~modes (middle panel).
We also show the amplitude spectrum in \echelle\ format (bottom panel) after first removing low-frequency variations (see Sec.~\ref{sec:xi2-cen}).
}
\label{fig:xi2-cen}
\end{figure}

The light curve for $\xi^2$~Cen (HD~113791; B3) shows both g and p modes (Fig.~\ref{fig:xi2-cen}).  The g~modes appear to have a regular structure that has a similar \firstnew{period} spacing to other SPB stars \citep{Papics++2017,Aerts2021,Szewczuk++2021}.
Remarkably, we also see a regular pattern of p modes in $\xi^2$~Cen (top panel of Fig.~\ref{fig:xi2-cen}).  Such regularity is common in solar-like oscillators \citep{Chaplin+Miglio2013} and also occurs in young \dsct\ stars \citep{Bedding++2020}, but is very unusual in \bcep\ stars.  One exception is V1449~Aql (HD~180642), for which CoRoT observations showed evidence for a large frequency spacing of $\Dnu = 2.3\,\cd$ \citep{Belkacem++2009}.  In $\xi^2$~Cen, the spacing between the peaks yields a large frequency spacing of $\Dnu = 3.46\,\cd$.  Using the standard scaling relation that \Dnu\ scales approximately with the square root of density \citep[e.g.][]{Aerts++2010-book}, this implies a mean stellar density of 0.087 in solar units.  This is in reasonable agreement with the rough estimate of 0.073 that we made in Sec.~\ref{sec:discussion}.  It is also interesting that the p~modes peaks are broadened, possibly indicating damping similar to that seen in V1449~Aql \citep{Belkacem++2009,Belkacem++2010}. 
The bottom panel of Fig.~\ref{fig:xi2-cen} shows the amplitude spectrum in \echelle\ format, after first removing low-frequency variations with a high-pass filter.  The pattern with two clear ridges is very reminiscent of those seen in young \dsct\ stars \citep{Bedding++2020}.  This star is clearly worthy of further study.

\subsection{HD~132094: a short-period heartbeat star}
\label{sec:hd-132094}

\begin{figure}
\centering
    \includegraphics[width=\columnwidth]{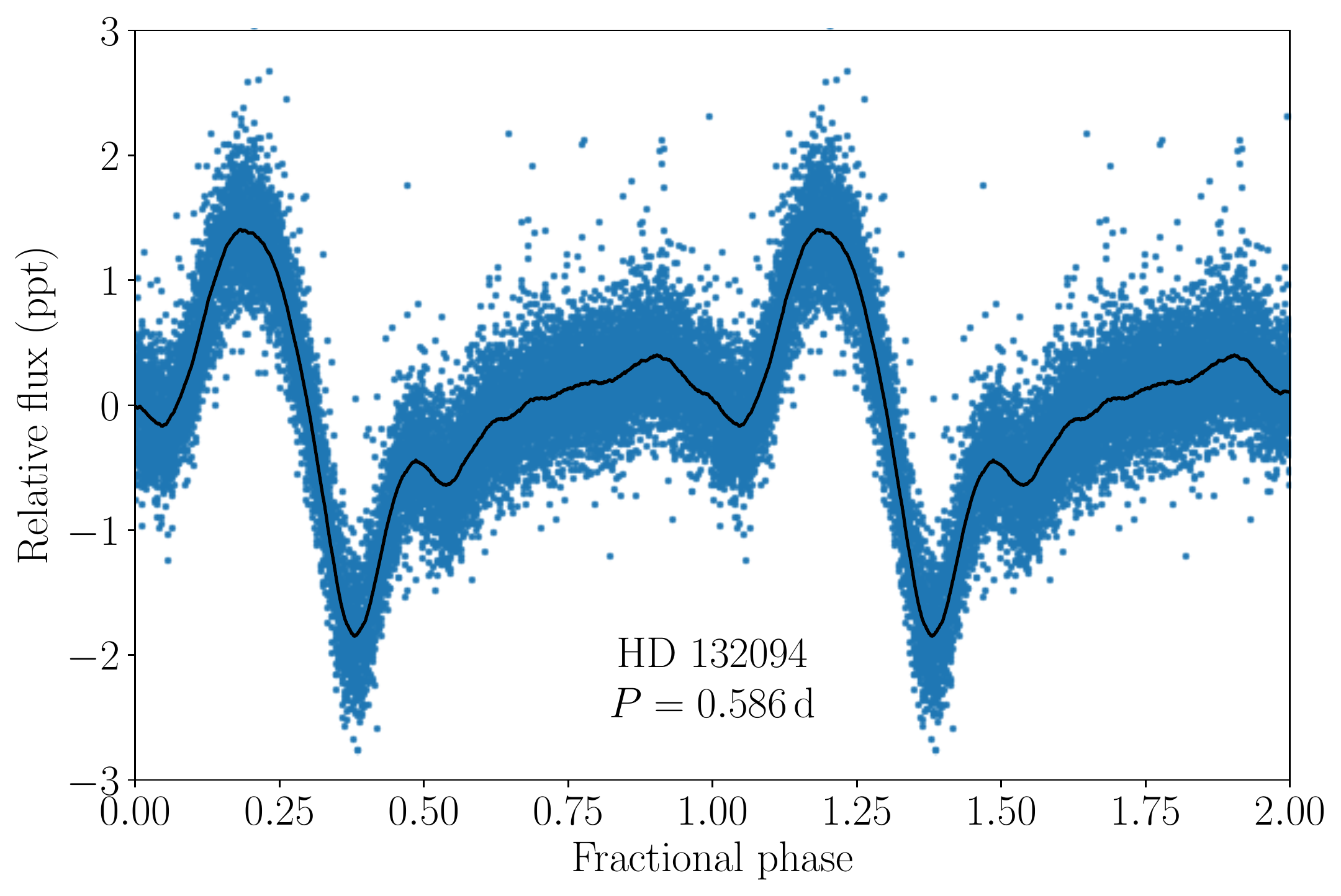}
\caption{Light curve of the heartbeat star HD~132094, folded at the orbital period.}
\label{fig:heartbeat}
\end{figure}

The star HD~132094 (spectral type B9) was not previously known to be a binary.  However, its \tess\ light curve (Fig.~\ref{fig:heartbeat}) shows it is a member of the class of eccentric binaries with tidal distortions (so-called heartbeat stars) that were discovered in large numbers by the \kepler\ Mission \citep[e.g.,][]{Thompson++2012,Kirk++2016}.  The period is \firstnew{$0.5865 \pm 0.0003$\,d}, which appears to be the shortest among previously published heartbeat stars \citep{Kirk++2016,Kolaczek-Szymanski++2021}.  The phase-folded light curve shows evidence for tidally-forced pulsations (not included in Table~\ref{tab:sample}) and this star certainly deserves further study.


\subsection{\texorpdfstring{$\tau$}{tau}~Lib: a heartbeat star with pulsations}
\label{sec:tau-lib}

\begin{figure}
\centering
    \includegraphics[width=\columnwidth]{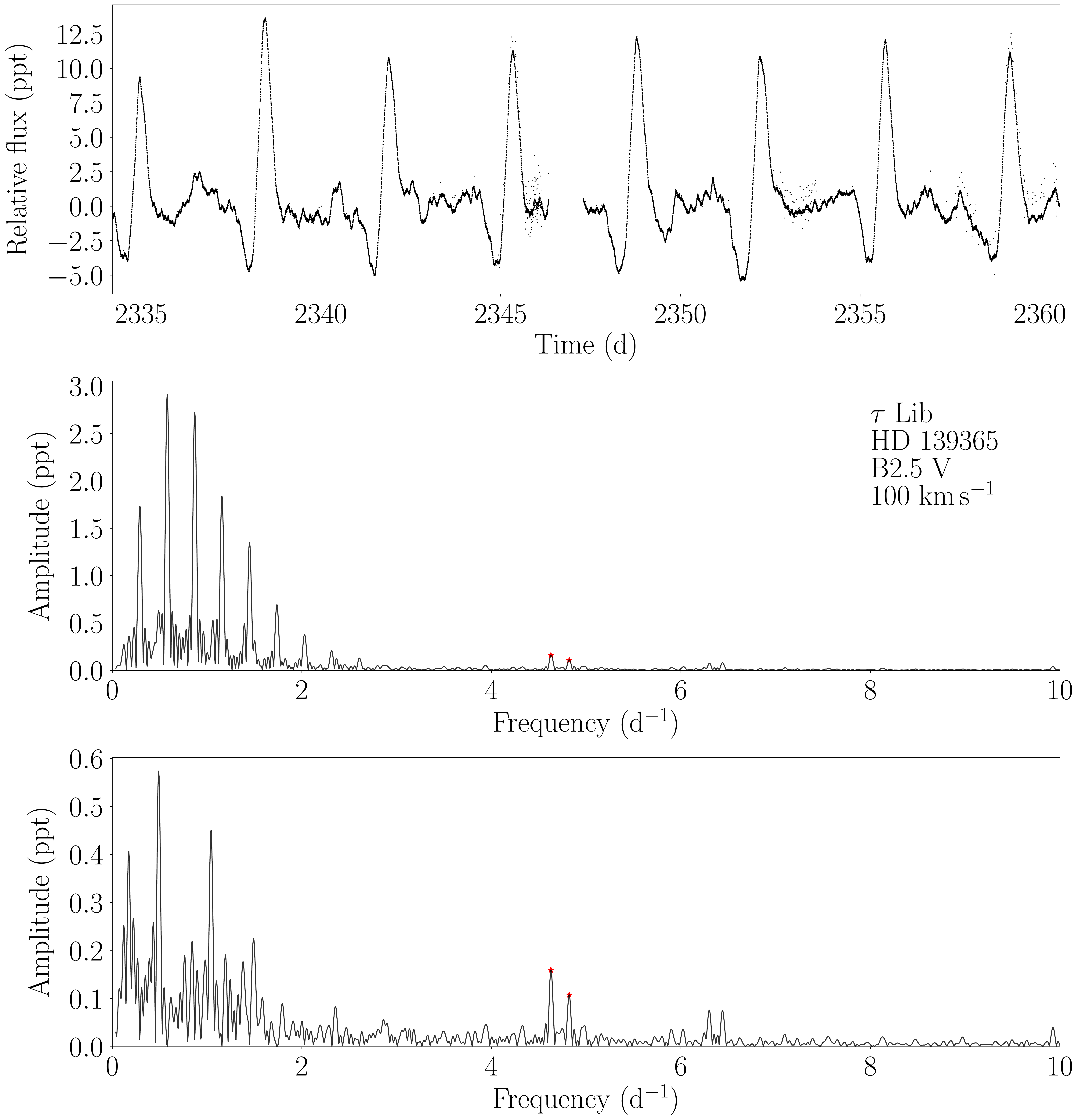}
\caption{\tess\ Light curve of $\tau$\ Lib.The top panel is original light curve, the middle panel is the Fourier Transform of the original light curve, and the bottom panel is the Fourier Transform of the light curve after removing the first 9 harmonics of the orbital frequency.}
\label{fig:tau-lib}
\end{figure}

The star $\tau$~Lib (HD~139365; B2.5) is known to be a spectroscopic binary, and photometry with the BRITE mission showed it to be a heartbeat star with a period of 3.41\,d \citep{Pigulski++2018-tau-Lib}. This is confirmed by the \tess\ light curve from Sector~38, which gives an orbital frequency of 0.290\,\cd\ and also shows clear pulsations (Fig.~\ref{fig:tau-lib}).  The two strongest pulsation peaks are listed in Table~\ref{tab:sample}, and we note that $f_1$ coincides with 16 times the orbital frequency.  


\subsection{\texorpdfstring{$\pi$}{pi}~Lup: an eclipsing binary with pulsations}
\label{sec:pi-lup}

\begin{figure}
\centering
    \includegraphics[width=\columnwidth]{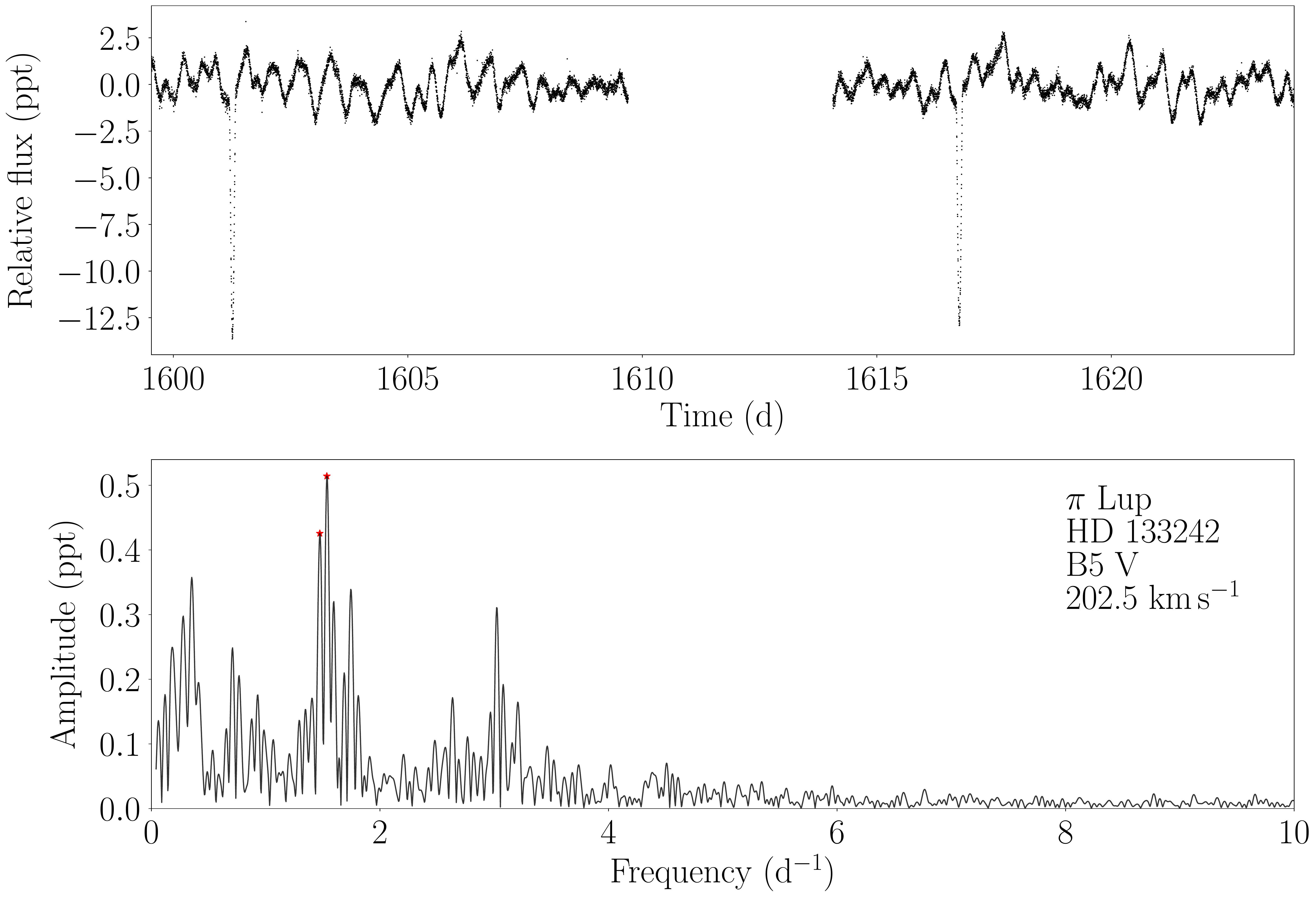}
\caption{\tess\ observations of $\pi$~Lup (B5\,V) from Sector~11.  The light curve (upper panel) shows it to be an eclipsing binary with pulsations.  The amplitude spectrum after removing the eclipses (lower panel) shows pulsations. 
}
\label{fig:pi-lup}
\end{figure}

$\pi$~Lup (HR~5605; HD~133242; B5\,V) is a multiple system with at least four components \citep{Nitschelm2004} but not previously known to be eclipsing.  The \tess\ light curve (Fig.~\ref{fig:pi-lup}) shows two narrow eclipses with depths of 1\% and separated by 15.50\,d.  The amplitude spectrum after removing the eclipses (shown in Fig.~\ref{fig:pi-lup} \new{and in the on-line supplementary material}) shows peaks at 1.5\,\cd\ that are consistent with SPB pulsations, or possibly r~modes. 


\subsection{HR~5846: an eclipsing binary with tidally-forced pulsations}
\label{sec:hr-5846}

\begin{figure}
\centering
    \includegraphics[width=\columnwidth]{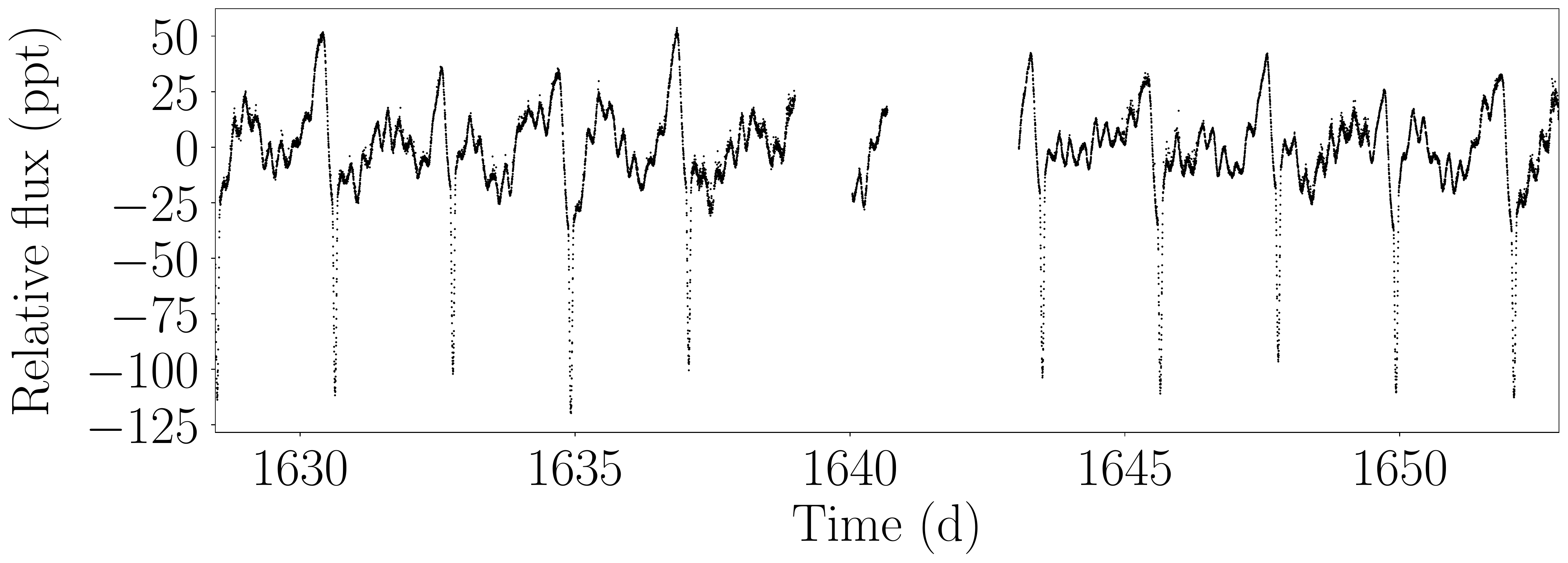}
\caption{\tess\ light curve of the star HR~5846 (A0\,V), showing eclipses and tidally forced pulsations. 
}
\label{fig:hr-5846}
\end{figure}

HR~5846 (HD~140285; A0\,V).  This star not in our sample because it does not have B spectral type but we mention it here because of its interesting light curve.  The \tess\ data (Fig.~\ref{fig:hr-5846}) shows detached eclipses (period 2.1 d) as well as pulsations at exactly 10 times the orbital frequency.  These are presumably tidally forced g modes and have an unusually large amplitude (almost 1\%).  Note that x-ray emission from HR~5846 was reported by \citet{Motch++1997}.

\subsection{Other eclipsing and close binaries}
\label{sec:eclipsing-binaries}

Light curves for the following stars are shown in the on-line supplementary material:
\begin{itemize}

\item $\mu^1$~Sco (HD~151890; B1.5 +B) is known to be a semi-detached eclipsing binary with a period of 1.45\,d \citep{vanAntwerpen+Moon2010}.  This is confirmed by the \tess\ light curve, which shows no sign of pulsations after prewhitening with the first 12 harmonics of the orbital frequency.


\item V964~Cen (HD~115823; B5) is catalogued as a short-period detached eclipsing binary \citep{Samus++2017} but the \tess\ light curve clearly shows it is actually an SPB star.

\item V831~Cen (HD~114529; B7) is known to be a close binary (ellipsoidal variable) with a period of 0.642\,d \citep{Budding++2010}, which is confirmed by the \tess\ light curve.

\item GG~Lup (HD~135876; B7) is known to be a detached eclipsing binary with a period of 1.85\,d \citep[e.g.,][]{Budding++2015}. This is confirmed by the \tess\ light curve, which also shows low-amplitude variability outside of the eclipses (around 1.6\,\cd) that may indicate tidally-forced pulsations.  

\item HR~4913 (HD~112409; V945~Cen; B8/9\,V) is known to be a close binary (ellipsoidal variable) with a period 0.65\,d \citep{Harmanec++2010}, which is confirmed by the \tess\ light curve.

\end{itemize}


\subsection{Other stars}
These stars are ordered by spectral type:
\begin{itemize}
\item $\tau$~Sco (HD~149438; B0\,V) is not catalogued as a \bcep\ star and no line profile variations have been reported \citep{Telting++2006}.
The TESS light curve shows no evidence for variability and we can rule out \bcep\ pulsations down to a level of about 50\,ppm.

\item $\beta$~Cru (HD~111123; B0.5\,III) is known to be a multi-periodic \bcep\ star, which is confirmed by the \tess\ light curve.  \cite{Aerts++1998} found it to be a spectroscopic binary in an eccentric 5-yr orbit, in which the primary (B0.5\,III) is the \bcep\ variable and the secondary is a B2\,V star.  The strongest modes in the \tess\ data have frequencies of 5.230, 5.487, 5.969, and 3.529\,\cd, which agree with those found by \citet{Cuypers++2002}
using photometry spanning 17\,d from the star tracker on the {\em WIRE} satellite.  
The \tess\ light curve from Sector~11 for this star has been analysed in detail by \citet{cotton++2022}.

\item $\alpha$~Lup (HD~129056; B1.5\,Vn) is known to be a single-mode \bcep\ star \citep{Lampens+Goossens1982,Mathias++1994,Nardetto++2013}.  The \tess\ light curve confirms the mode at 3.848\,\cd\ (7.0\,ppt), and subtracting this mode shows much weaker peaks (0.6--0.7\,ppt) at 3.318, 3.916 and 3.976\,\cd.

\item $\nu$~Cen (HD~120307; B2)
This system is a spectroscopic binary with an orbital period of 2.625\,d.
\firstnew{Using photometry from the SMEI and BRITE spacecraft, \citet{Jerzykiewicz++2021} found variations with this period due to the reflection effect.}
The TESS 2-minute light curve \firstnew{confirms} these variations.  After removing them, we find some irregular variability but no evidence for coherent pulsations.

\item $\gamma$~Lup (HD~138690; B2\,IV; $V = 2.77$)
This is a visual binary with nearly equal components separated by about half an arcsecond \citep{Herschel1847}. 
One of the components is a spectroscopic binary with an orbital period of 2.850\,d.
It was not previously reported as a \bcep\ star.
\firstnew{Using photometry from the SMEI and BRITE spacecraft, \citet{Jerzykiewicz++2021} found variations at the orbital period due to the reflection effect.}
The TESS 2-minute light curve \firstnew{confirms} these variations.  After removing them, the light curve appears to show several low-amplitude peaks in the range 1.5 to 15\,\cd\ that may indicate SPB and \bcep\ pulsations, but these are not listed in the table.

\end{itemize}






\section{Conclusions}

We have examined the light curves of 119 B-type stars in the Sco--Cen Association that have been observed by \tess.  Pulsations are seen in 81 stars (68\%), covering the whole spectral range in the H--R diagram.  In particular, we confirm the presence of low-frequency pulsations in stars whose effective temperatures lie between those  normally associated with SPB and \dsct\ stars.  By taking the stellar densities into account, we conclude that these cannot be p~modes and confirm previous suggestions that these are probably rapidly-rotating SPB stars. We also confirm that they follow two period--luminosity relations that are consistent with prograde sectoral g~modes that are dipole ($l=m=1$) and quadrupole ($l=m=2$), respectively.  

One of the stars ($\xi^2$~Cen; Sec.~\ref{sec:xi2-cen}) is a hybrid pulsator that shows regular spacings in both g~and p~modes.  In particular, the p~modes show a remarkable regularity that is reminiscent of solar-like oscillations.
Our sample also includes several interesting binaries, including 
a very short-period heartbeat star (HD~132094; Sec.~\ref{sec:hd-132094}), 
a previously unknown eclipsing binary ($\pi$~Lup; Sec~\ref{sec:pi-lup}), and
an eclipsing binary with high-amplitude tidally driven pulsations (HR~5846; Sec.~\ref{sec:hr-5846}).
Overall, the results show the power of \tess\ for studying variability in bright stars.

\section*{Acknowledgements}

We thank Jim Fuller, Mike Ireland, \firstnew{Andrzej Pigulski,} Aaron Rizzuto and Rich Townsend for helpful discussions.
We gratefully acknowledge support from the Australian Research Council through Discovery Project DP210103119, and
from the Danish National Research Foundation (Grant DNRF106) through its
funding for the Stellar Astrophysics Center (SAC).
%
We are grateful to the entire \tess\ team for providing the data used in this paper. 
This work made use of several publicly available {\tt python} packages: 
{\tt astropy} \citep{astropy:2013,astropy:2018}, 
{\tt echelle} \citep{Hey+Ball2019},
{\tt lightkurve} \citep{lightkurve2018},
{\tt matplotlib} \citep{matplotlib2007}, 
{\tt numpy} \citep{numpy2020}, 
and 
{\tt scipy} \citep{scipy2020}.

\section*{Data Availability}

The data underlying this article are available at the MAST Portal (Barbara A. Mikulski Archive for Space Telescopes), at \url{https://mast.stsci.edu/portal/Mashup/Clients/Mast/Portal.html}

\input{output.bbl} 


\clearpage
\onecolumn
\begin{longtable}{rrrrlrrrrrrr}
\caption{Sample of B stars in Sco--Cen that have been observed by \tess\ (ordered by spectral type).\label{tab:sample}}\\
Name                  & HD       & HIP     & V   & Sp. type& $\log(\Teff/{\rm K})$ & $\log(L/\Lsun)$ & $v\sin i$  & $f_1$     & $f_2$     & $f_3$    &$f_4$ \\
      &     &   &   &   &  &   &(\kms)   & (\cd) & (\cd) & (\cd) & (\cd) \\
\noalign{\smallskip}\hline\noalign{\smallskip}
\endfirsthead
\caption[]{(continued from previous page)}\\
Name                  & HD       & HIP     & V   & Sp. type& $\log(\Teff/{\rm K})$ & $\log(L/\Lsun)$ & $v\sin i$  & $f_1$     & $f_2$     & $f_3$    &$f_4$ \\
      &     &   &   &   &  &   &(\kms)   & (\cd) & (\cd) & (\cd) & (\cd) \\
\noalign{\smallskip}\hline\noalign{\smallskip}
\endhead
\input{mytable.tex}
\end{longtable}

\clearpage
\twocolumn

\bsp	
\label{lastpage}
\end{document}

%% file: mytable.tex
$\tau$\ Sco  & 149438 & 81266 & 2.81 &  B0 V                & 4.50 & $\  4.31\pm0.07$\   &    8.0 &   --- &   --- &  --- &  --- \\      $\beta$\ Cru  & 111123 & 62434 & 1.25 &    B0.5 III          & 4.46 & $\  4.41\pm0.07$\   &   39.5 &  5.23 &  5.48 &  --- &  --- \\     $\alpha$\ Cru  & 108248 & 60718 & 1.28 &    B0.5 IV           & 4.46 & $\  4.52\pm0.04$\   &  124.0 &  3.05 &  9.47 &  --- &  --- \\      $\beta$\ Cen  & 122451 & 68702 & 0.58 &              B1 III  & 4.42 & $\  4.55\pm0.05$\   &  133.0 &  5.55 &  6.41 &  --- &  --- \\     $\delta$\ Lup  & 136298 & 75141 & 3.19 &             B1.5 IV  & 4.39 & $\  3.94\pm0.06$\   &  232.5 &  5.90 &  2.54 &  --- &  --- \\      $\mu^1$\ Sco  & 151890 & 82514 & 2.98 &  B1.5 IV + B         & 4.40 & $\  4.05\pm0.12$\   &  180.0 &   --- &   --- &  --- &  --- \\     $\alpha$\ Lup  & 129056 & 71860 & 2.29 &             B1.5 Vn  & 4.39 & $\  4.26\pm0.02$\   &    9.5 &  3.85 &   --- &  --- &  --- \\           HR 6143  & 148703 & 80911 & 4.23 &  B2 III-IV           & 4.31 & $\  3.55\pm0.03$\   &   81.5 &  4.46 &  4.67 & 1.97 &  --- \\     $\tau^1$\ Lup  & 126341 & 70574 & 4.55 &    B2 IV             & 4.31 & $\  3.71\pm0.05$\   &    7.5 &  5.64 &  5.58 &  --- &  --- \\     $\alpha$\ Mus  & 109668 & 61585 & 2.65 &    B2 IV             & 4.31 & $\  3.63\pm0.01$\   &  114.0 &  7.17 &  5.48 &  --- &  --- \\      $\beta$\ Lup  & 132058 & 73273 & 2.68 &    B2 IV             & 4.31 & $\  3.79\pm0.02$\   &  130.0 &   --- &   --- & 0.53 & 0.58 \\     $\delta$\ Cru  & 106490 & 59747 & 2.75 &    B2 IV             & 4.31 & $\  3.91\pm0.09$\   &  214.0 &  2.80 &  2.58 & 0.88 & 1.42 \\     $\kappa$\ Cen  & 132200 & 73334 & 3.11 &   B2 IV              & 4.31 & $\  3.62\pm0.06$\   &   20.0 &   --- &   --- &  --- &  --- \\        $\nu$\ Cen  & 120307 & 67464 & 3.39 &   B2 IV              & 4.31 & $\  3.55\pm0.04$\   &   92.0 &   --- &   --- &  --- &  --- \\     $\gamma$\ Lup  & 138690 & 76297 & 2.76 &   B2 IV              & 4.31 & $\  3.84\pm0.06$\   &  276.5 &   --- &   --- &  --- &  --- \\      $\mu^2$\ Sco  & 151985 & 82545 & 3.54 &  B2 IV               & 4.31 & $\  3.80\pm0.04$\   &   49.0 &  2.82 &  3.51 &  --- &  --- \\ $\upsilon^1$\ Cen  & 121790 & 68282 & 3.87 &   B2 IV/V            & 4.31 & $\  3.41\pm0.02$\   &  124.0 &  2.98 &   --- &  --- &  --- \\       $\rho$\ Sco  & 142669 & 78104 & 3.86 &  B2 IV/V             & 4.31 & $\  3.45\pm0.04$\   &  117.0 &  3.81 & 12.37 & 0.40 & 1.17 \\        $\mu$\ Cen  & 120324 & 67472 & 3.43 &     B2 IV/Vne        & 4.31 & $\  3.62\pm0.06$\   &  192.5 &  3.35 &  3.12 &  --- &  --- \\       $\chi$\ Cen  & 122980 & 68862 & 4.34 &   B2 V               & 4.31 & $\  3.36\pm0.03$\   &   10.0 &  2.46 &  4.16 &  --- &  --- \\       $\phi$\ Cen  & 121743 & 68245 & 3.80 &   B2 V               & 4.31 & $\  3.50\pm0.03$\   &  107.5 &  5.10 &  5.16 & 0.88 & 0.98 \\     $\sigma$\ Cen  & 108483 & 60823 & 3.91 &   B2 V               & 4.31 & $\  3.36\pm0.03$\   &  169.0 &  2.97 &  3.27 & 1.34 & 1.27 \\          V761 Cen  & 125823 & 70300 & 4.42 &  B2 V                & 4.31 & $\  3.11\pm0.02$\   &    5.0 &   --- &   --- & 1.18 &  --- \\            13 Sco  & 145482 & 79404 & 4.57 &  B2 V                & 4.31 & $\  3.23\pm0.02$\   &  198.5 &   --- &   --- & 1.89 &  --- \\       $\eta$\ Lup  & 143118 & 78384 & 3.41 &  B2.5 IV             & 4.27 & $\  3.49\pm0.05$\   &  241.0 &  6.57 &  6.06 &  --- &  --- \\           HR 5471  & 129116 & 71865 & 4.00 &  B2.5 V              & 4.27 & $\  3.01\pm0.03$\   &  198.5 &  5.43 &   --- & 1.23 & 2.28 \\       $\tau$\ Lib  & 139365 & 76600 & 3.64 &  B2.5 V              & 4.27 & $\  3.26\pm0.02$\   &  100.0 &  4.63 &  4.82 &  --- &  --- \\     $\theta$\ Lup  & 144294 & 78918 & 4.20 &             B2.5 Vn  & 4.27 & $\  3.20\pm0.03$\   &  330.5 &  3.75 &  3.69 & 2.17 & 1.85 \\           HR 4848  & 110956 & 62327 & 4.62 &    B2/3 V            & 4.27 & $\  2.79\pm0.02$\   &   11.0 &   --- &   --- & 1.24 & 1.94 \\      $\zeta$\ Cru  & 106983 & 60009 & 4.05 &    B2/3 V            & 4.27 & $\  3.06\pm0.02$\   &  151.5 &  2.61 &  5.28 & 0.47 & 2.22 \\       $\eta$\ Cen  & 127972 & 71352 & 2.31 &                B2Ve  & 4.31 & $\  3.89\pm0.02$\   &  305.0 &   --- &   --- &  --- &  --- \\           HR 5035  & 116087 & 65271 & 4.51 &                B3 V  & 4.23 & $\  2.83\pm0.01$\   &  223.0 &  4.15 &  4.03 & 1.57 & 2.14 \\      $\xi^2$\ Cen  & 113791 & 64004 & 4.27 &   B3 V               & 4.23 & $\  3.16\pm0.02$\   &   32.0 & 17.23 & 15.59 & 0.77 & 0.92 \\            KT Lup  & 138769 & 76371 & 4.71 &   B3 V               & 4.23 & $\  2.97\pm0.02$\   &  106.0 &   --- &   --- &  --- &  --- \\       $\rho$\ Cen  & 105937 & 59449 & 3.96 &   B3 V               & 4.23 & $\  2.75\pm0.03$\   &  147.0 &  3.70 &   --- & 0.61 & 1.12 \\           HR 6214  & 150742 & 81972 & 5.63 &   B3 V               & 4.23 & $\  2.76\pm0.01$\   &  167.0 &   --- &   --- &  --- &  --- \\           HR 5595  & 132955 & 73624 & 5.44 &  B3 V                & 4.23 & $\  2.68\pm0.02$\   &   10.0 &   --- &   --- &  --- &  --- \\           HR 4732  & 108257 & 60710 & 4.81 &   B3 V(n)            & 4.23 & $\  2.87\pm0.02$\   &  298.0 &  4.81 &  4.75 & 0.61 & 2.17 \\    $\lambda$\ Cru  & 112078 & 63007 & 4.60 &        B3 Vne        & 4.23 & $\  2.80\pm0.01$\   &  341.0 &  3.23 &  3.57 &  --- &  --- \\       $\rho$\ Lup  & 128345 & 71536 & 4.05 &   B3/4 V             & 4.22 & $\  2.95\pm0.02$\   &  240.0 &   --- &   --- & 2.25 & 2.13 \\          V863 Cen  & 105382 & 59173 & 4.47 &   B3/5 III           & 4.22 & $\  2.79\pm0.02$\   &   70.0 &   --- &   --- &  --- &  --- \\             3 Cen  & 120709 & 67669 & 4.52 &  B4 III              & 4.22 & $\  2.71\pm0.01$\   &    0.0 &   --- &   --- &  --- &  --- \\           HR 4549  & 103079 & 57851 & 4.99 &    B4 V              & 4.22 & $\  2.64\pm0.01$\   &   49.0 &   --- &   --- &  --- &  --- \\           HR 4940  & 113703 & 63945 & 4.69 &   B4 V               & 4.22 & $\  2.81\pm0.02$\   &  216.0 &  3.17 &  3.11 & 1.76 & 1.59 \\          V964 Cen  & 115823 & 65112 & 5.44 &   B5 III/IV          & 4.20 & $\  2.42\pm0.01$\   &   41.0 &   --- &   --- & 1.29 & 1.36 \\         $o$\ Lup  & 130807 & 72683 & 4.31 &   B5 IV              & 4.20 & $\  2.91\pm0.06$\   &   51.0 &   --- &   --- & 1.10 & 0.34 \\        $\pi$\ Lup  & 133242 & 73807 & 4.58 &   B5 V               & 4.20 & $\  2.88\pm0.07$\   &  202.5 &   --- &   --- & 1.54 & 1.47 \\     $\psi^2$\ Lup  & 140008 & 76945 & 4.72 &  B5 V                & 4.20 & $\  2.78\pm0.02$\   &   46.0 &   --- &   --- & 1.17 & 1.07 \\           HR 5736  & 137432 & 75647 & 5.45 &  B5 V                & 4.20 & $\  2.57\pm0.01$\   &  130.0 &  4.48 &  6.51 & 2.13 & 1.15 \\        $\pi$\ Cen  &  98718 & 55425 & 4.08 &    B5 Vn             & 4.20 & $\  2.52\pm0.06$\   &  351.5 &  5.21 &  2.94 & 2.27 & 2.33 \\      $\mu^2$\ Cru  & 112091 & 63005 & 5.20 &              B5 Vne  & 4.20 & $\  2.53\pm0.01$\   &  124.0 &  2.97 &   --- & 0.62 & 1.07 \\          V795 Cen  & 124367 & 69618 & 5.07 &              B5 Vne  & 4.20 & $\  2.65\pm0.02$\   &  301.0 &  2.77 &  2.72 & 1.87 &  --- \\           HR 5967  & 143699 & 78655 & 4.89 &  B5/7 III/IV         & 4.16 & $\  2.66\pm0.01$\   &  175.0 &  5.93 &   --- & 1.12 & 1.06 \\           HR 5625  & 133937 & 74100 & 5.82 &   B5/7 V             & 4.16 & $\  2.35\pm0.01$\   &  350.0 &  5.92 &  5.86 &  --- &  --- \\           HR 6209  & 150591 & 81914 & 6.12 &   B6/7 V             & 4.15 & $\  2.38\pm0.01$\   &  271.0 &  4.54 &  6.47 & 1.89 & 1.94 \\           HR 5753  & 138221 & 76048 & 6.49 &  B6/7 V              & 4.15 & $\  2.17\pm0.01$\   &  240.0 &  2.92 &  2.99 &  --- &  --- \\         V1019 Cen  & 131120 & 72800 & 5.01 &  B7 II/III           & 4.15 & $\  2.63\pm0.02$\   &   60.0 &   --- &   --- &  --- &  --- \\          V831 Cen  & 114529 & 64425 & 4.59 &    B7 V              & 4.15 & $\  2.57\pm0.01$\   &  216.0 &   --- &   --- &  --- &  --- \\            GG Lup  & 135876 & 74950 & 5.60 &   B7 V               & 4.15 & $\  2.44\pm0.01$\   &  129.0 &   --- &   --- &  --- &  --- \\           HR 5439  & 127971 & 71353 & 5.87 &   B7 V               & 4.15 & $\  2.05\pm0.02$\   &  191.0 &  5.42 &  3.04 & 2.25 & 2.18 \\                --- & 147001 & 80142 & 6.50 &   B7 V               & 4.15 & $\  2.22\pm0.01$\   &    --- &   --- &   --- &  --- &  --- \\           HR 5910  & 142250 & 77900 & 6.10 &  B7 V                & 4.15 & $\  2.23\pm0.01$\   &   15.0 &  2.54 &  2.50 &  --- &  --- \\           HR 4706  & 107696 & 60379 & 5.37 &          B7 Vn       & 4.15 & $\  2.21\pm0.01$\   &  321.0 &  3.32 &   --- &  --- &  --- \\           HR 4834  & 110506 & 62058 & 5.98 &    B7/8 V            & 4.12 & $\  2.06\pm0.01$\   &  178.0 &  2.71 &  2.76 & 1.28 &  --- \\           HR 4879  & 111774 & 62786 & 5.97 &  B7/8 V              & 4.12 & $\  2.17\pm0.01$\   &   64.0 &   --- &   --- & 1.92 & 1.82 \\            52 Hya  & 126769 & 70753 & 4.97 &  B7/8 V              & 4.12 & $\  2.53\pm0.01$\   &  199.5 &  2.98 &  2.92 & 0.45 & 0.51 \\           HR 4290  &  95324 & 53701 & 6.18 &    B8 IV             & 4.09 & $\  1.88\pm0.01$\   &   55.0 &  5.52 &   --- &  --- &  --- \\           HR 4089  &  90264 & 50847 & 4.99 &                B8 V  & 4.09 & $\  2.44\pm0.01$\   &   11.0 &   --- &   --- &  --- &  --- \\           HR 4355  &  97583 & 54767 & 5.22 &    B8 V              & 4.09 & $\  2.10\pm0.01$\   &  178.0 &   --- &   --- & 1.90 & 1.85 \\            FH Mus  & 110020 & 61796 & 6.25 &    B8 V              & 4.09 & $\  1.82\pm0.01$\   &  234.0 &  2.88 &  2.93 &  --- &  --- \\           HR 5121  & 118354 & 66454 & 5.89 &   B8 V               & 4.09 & $\  2.17\pm0.02$\   &  253.0 &   --- &   --- & 1.33 & 1.38 \\                --- & 126135 & 70455 & 6.96 &   B8 V               & 4.09 & $\  1.94\pm0.01$\   &    --- &   --- &   --- &  --- &  --- \\                --- & 142256 & 77968 & 6.96 &   B8 V               & 4.09 & $\  1.95\pm0.01$\   &    --- &  3.75 &  3.81 &  --- &  --- \\           HR 5207  & 120642 & 67703 & 5.26 &   B8 V               & 4.09 & $\  2.05\pm0.01$\   &    --- &  6.90 &  3.51 &  --- &  --- \\           HR 6100  & 147628 & 80390 & 5.41 &  B8 V                & 4.09 & $\  2.28\pm0.01$\   &  160.0 &  2.54 &  2.59 &  --- &  --- \\           HR 5579  & 132238 & 73341 & 6.47 &  B8 V                & 4.09 & $\  2.02\pm0.01$\   &  234.0 &  3.13 &  6.75 &  --- &  --- \\           HR 5860  & 140784 & 77286 & 5.60 &  B8 V                & 4.09 & $\  2.15\pm0.01$\   &  364.0 &  3.05 &  6.08 &  --- &  --- \\           HR 5449  & 128207 & 71453 & 5.74 &  B8 V                & 4.09 & $\  2.19\pm0.01$\   &    --- &  6.34 &  6.40 & 0.50 & 0.57 \\           HR 6211  & 150638 & 81891 & 6.45 &  B8 V                & 4.09 & $\  2.00\pm0.01$\   &  166.0 &   --- &   --- &  --- &  --- \\        $\mu$\ Lup  & 135734 & 74911 & 4.94 &  B8 Ve               & 4.09 & $\  2.26\pm0.06$\   &  308.0 & 33.16 & 33.21 &  --- &  --- \\           HR 4221  &  93563 & 52742 & 5.23 &    B8/9 III          & 4.06 & $\  2.55\pm0.02$\   &  249.0 &   --- &   --- & 1.15 & 1.10 \\           HR 4913  & 112409 & 63210 & 5.16 &   B8/9 V             & 4.06 & $\  2.30\pm0.01$\   &  123.0 &   --- &   --- &  --- &  --- \\                --- & 104080 & 58452 & 6.34 &   B8/9 V             & 4.06 & $\  1.86\pm0.01$\   &    --- &   --- &   --- &  --- &  --- \\                --- & 128819 & 71724 & 6.65 &   B8/9 V             & 4.06 & $\  1.84\pm0.01$\   &    --- &   --- &   --- &  --- &  --- \\           HR 4951  & 113902 & 64053 & 5.70 &   B8/9 V             & 4.06 & $\  1.84\pm0.01$\   &  249.0 &   --- &   --- &  --- &  --- \\           HR 4748  & 108541 & 60855 & 5.44 &  B8/9 V              & 4.06 & $\  2.32\pm0.02$\   &  191.0 &   --- &   --- & 0.96 & 0.32 \\                --- & 136482 & 75210 & 6.64 &  B8/9 V              & 4.06 & $\  1.80\pm0.01$\   &    --- &  3.57 &  3.62 &  --- &  --- \\                --- & 143927 & 78754 & 7.06 &  B8/9 V              & 4.06 & $\  1.81\pm0.02$\   &    --- &   --- &   --- &  --- &  --- \\           HR 5790  & 138923 & 76395 & 6.25 &  B8/9 V              & 4.06 & $\  1.90\pm0.01$\   &  271.0 &  3.32 &  3.37 &  --- &  --- \\                --- & 151109 & 82154 & 7.00 &  B9 IV/V             & 4.03 & $\  1.76\pm0.01$\   &    --- &  3.65 &  3.61 &  --- &  --- \\           HR 4692  & 107301 & 60183 & 6.19 &    B9 V              & 4.03 & $\  1.57\pm0.01$\   &  209.0 &   --- &   --- &  --- &  --- \\           HR 4597  & 104600 & 58720 & 5.88 &    B9 V              & 4.03 & $\  1.75\pm0.01$\   &  214.0 &  3.33 &  3.38 &  --- &  --- \\           HR 4832  & 110461 & 62026 & 6.06 &    B9 V              & 4.03 & $\  1.71\pm0.01$\   &  246.0 &  3.53 &  3.58 &  --- &  --- \\                --- & 115583 & 65021 & 7.25 &    B9 V              & 4.03 & $\  1.64\pm0.01$\   &    --- &  3.38 &  3.24 &  --- &  --- \\           HR 5230  & 121190 & 67973 & 5.71 &   B9 V               & 4.03 & $\  1.95\pm0.01$\   &  166.0 &  2.62 &  2.68 &  --- &  --- \\                --- & 115988 & 65178 & 6.66 &   B9 V               & 4.03 & $\  1.62\pm0.01$\   &    --- &  2.85 &   --- &  --- &  --- \\           HR 5294  & 123445 & 69113 & 6.17 &   B9 V               & 4.03 & $\  2.01\pm0.02$\   &   66.0 &  3.71 &  3.77 & 0.63 & 0.68 \\           HR 5773  & 138564 & 76234 & 6.35 &  B9 V                & 4.03 & $\  1.53\pm0.01$\   &  154.0 &   --- &   --- &  --- &  --- \\            12 Sco  & 145483 & 79399 & 5.80 &  B9 V                & 4.03 & $\  1.76\pm0.02$\   &  204.0 &   --- &   --- & 1.66 & 1.60 \\           HR 5805  & 139233 & 76591 & 6.58 &  B9 V                & 4.03 & $\  1.77\pm0.01$\   &  209.0 &   --- &   --- &  --- &  --- \\           HR 5400  & 126475 & 70626 & 6.35 &  B9 V                & 4.03 & $\  1.83\pm0.01$\   &  234.0 &  3.32 &  3.37 &  --- &  --- \\                --- & 132094 & 73266 & 7.25 &  B9 V                & 4.03 & $\  1.64\pm0.01$\   &    --- &   --- &   --- &  --- &  --- \\                --- & 141327 & 77523 & 7.46 &  B9 V                & 4.03 & $\  1.73\pm0.01$\   &    --- &   --- &   --- &  --- &  --- \\                --- & 144591 & 79044 & 6.74 &  B9 V                & 4.03 & $\  1.66\pm0.01$\   &    --- &   --- &   --- &  --- &  --- \\                --- & 149274 & 81208 & 6.64 &  B9 V                & 4.03 & $\  1.75\pm0.01$\   &    --- &   --- &   --- &  --- &  --- \\                --- & 151726 & 82430 & 7.24 &  B9 V                & 4.03 & $\  1.64\pm0.01$\   &    --- &   --- &   --- &  --- &  --- \\           HR 4874  & 111597 & 62683 & 4.89 &  B9 V                & 4.03 & $\  2.31\pm0.02$\   &    --- & 14.21 & 26.75 & 2.36 & 1.49 \\           HR 5869  & 141168 & 77562 & 5.78 &   B9 V(n)            & 4.03 & $\  1.73\pm0.01$\   &  308.0 &  3.99 &  7.52 &  --- &  --- \\           HR 5141  & 118991 & 66821 & 5.22 &        B9 Vn         & 4.03 & $\  1.88\pm0.01$\   &  321.0 &  3.20 &  2.63 &  --- &  --- \\                --- & 104900 & 58901 & 6.20 &        B9 Vn         & 4.03 & $\  1.73\pm0.01$\   &    --- &  3.79 &  3.67 &  --- &  --- \\           HR 4985  & 114772 & 64515 & 5.90 &   B9.5 V             & 4.02 & $\  1.79\pm0.07$\   &  249.0 &  2.61 &  3.64 &  --- &  --- \\                --- & 104839 & 58859 & 6.47 &   B9.5 V             & 4.02 & $\  1.62\pm0.01$\   &    --- &   --- &   --- &  --- &  --- \\                --- & 109195 & 61257 & 6.55 &   B9.5 V             & 4.02 & $\  1.63\pm0.01$\   &    --- &   --- &   --- &  --- &  --- \\                --- & 115470 & 64892 & 6.79 &   B9.5 V             & 4.02 & $\  1.49\pm0.01$\   &    --- &   --- &   --- &  --- &  --- \\                --- & 117484 & 65965 & 7.53 &   B9.5 V             & 4.02 & $\  1.42\pm0.01$\   &    --- &   --- &   --- &  --- &  --- \\                --- & 123247 & 69011 & 6.42 &   B9.5 V             & 4.02 & $\  1.46\pm0.01$\   &    --- &   --- &   --- &  --- &  --- \\                --- & 135454 & 74752 & 6.75 &   B9.5 V             & 4.02 & $\  1.58\pm0.01$\   &    --- &   --- &   --- &  --- &  --- \\                --- & 137919 & 75915 & 6.33 &   B9.5 V             & 4.02 & $\  1.91\pm0.01$\   &    --- &  2.82 &  2.88 &  --- &  --- \\                --- & 143022 & 78324 & 8.18 &   B9.5 V             & 4.02 & $\  1.22\pm0.01$\   &    --- &   --- &   --- & 2.18 & 2.24 \\